\nonstopmode
\documentclass[11pt,authoryear]{article}
\usepackage{graphicx}
\usepackage[paperheight=11.7in,paperwidth=8.3in]{geometry}
\usepackage{enumerate}
\usepackage{url}
\usepackage[english]{babel}
\usepackage{amssymb}
\usepackage{amsfonts}
\usepackage{color}
\usepackage{epstopdf}
\usepackage{dsfont}
\usepackage{bm}
\usepackage{authblk}
\usepackage[authoryear]{natbib}
\usepackage[safe]{tipa}

\title{The statistical analysis of acoustic phonetic data:\\
exploring differences between spoken Romance languages.
}
\author{Davide Pigoli\footnote{Address for correspondence: Davide Pigoli,
Statistical laboratory, Department of Pure Mathematics and
Mathematical Statistics, University of Cambridge, Wilberforce Road,
Cambridge, CB3 0WB, United Kingdom. Email:
dp497@cam.ac.uk}\hspace{1mm}} \affil{Statistical Laboratory, DPMMS,
University of Cambridge} \author{Pantelis Z. Hadjipantelis}
\affil{Department of Statistics, UC Davis} \author{John S.
Coleman}\affil{Phonetics Laboratory, University of Oxford}
\author{John A.D. Aston} \affil{Statistical Laboratory, DPMMS,
University of Cambridge}
\date{}

\begin{document}

\maketitle
\begin{abstract}
The historical and geographical spread from older to more modern languages has long been studied by examining textual changes and in terms of changes in phonetic transcriptions. However, it is more difficult to analyze language change from an acoustic point of view, although this is usually the dominant mode of transmission. We propose a novel analysis approach for acoustic phonetic data, where the aim will be to statistically model the acoustic properties of spoken words. We explore phonetic variation and change using a time-frequency representation, namely the log-spectrograms of speech recordings. We identify time and frequency covariance functions as a feature of the language; in contrast, mean spectrograms depend mostly on the particular word that has been uttered. We build models for the mean and covariances (taking into account the restrictions placed on the statistical analysis of such objects) and use these to define a phonetic transformation that models how an individual speaker would sound in a different language, allowing the exploration of phonetic differences between languages. Finally, we map back these transformations to the domain of sound recordings, allowing us to listen to the output of the statistical analysis. The proposed approach is demonstrated using recordings of the words corresponding to the numbers from ``one'' to ``ten'' as pronounced by speakers from five different Romance languages.
\end{abstract}

\section{Introduction}
\label{sec:intro}

Historical and comparative linguistics is the branch of linguistics
which studies languages' evolution and relationships. The idea that
languages develop historically by a process roughly similar to
biological evolution is now generally accepted; see, e.g., \citet{cavallisforza1997} and
\citet{Nakhleh2005}. \citet{Pagel2009} claims that genes and languages have similar evolutionary behaviour and
offers an extensive catalogue of analogies between biological and
linguistic evolution. This immediately gives rise to the notion of familial relationships between languages.

However, interest in language kinships is not by any means restricted to
linguistics. For example, the understanding of this evolutionary
process is helpful for anthropologists and geneticists, while
distances between languages are proxies for cultural differences and
communication difficulties and can be used as such in sociology
and economic models \citep[][]{Ginsburgh2011}. Moreover, the nature of the relationship between languages, and especially the way they are spoken, is a topic of widespread interest for its cultural relevance. We all have our own experience with learning and using different languages (and different varieties within each language) and the effort to find quantitative properties of speech can shed some light on the subject.

The first step in exploring the language ecosystem is to choose how to
analyse and measure the differences between languages. A language is a complex entity and its evolution can be considered
from many different points of view. The processes of change from one
language to another have long been studied by considering
textual and phonetic representation of the words \citep[see,
e.g.,][and references therein]{Davies98}. This focus on written forms reflects a general
normative approach towards languages: for cultural and historical
reasons, the way we think about them is
focused on the written expression of the words and their
``proper'' pronunciations. However, this is more a social artifact
than a reality of the population, as there is great variation within each
language depending on socio-economic and physiological attributes,
geography and other factors.

The focus of this work is on a more recent development in
quantitative linguistics: the study of acoustic phonetic variation,
i.e. change in the sounds associated with the pronunciations of words. On
the one hand, these provide a complementary way to consider the
difference between two languages which can be juxtaposed with the
differences measured using textual and phonetic representation. On the
other hand, it can be claimed that the acoustic expression of the
word is a more natural object of interest, textual and phonetic
transcriptions being only the representation used by linguists of
the normative (or more careful) pronunciations of words. However, the
use of speech recordings from actual speakers is not yet well
established in historical linguistics, due to the complexity of
speech as a data object, the theoretical challenges on how to
deal with the variability within and between languages and the difficulties (or impossibility) of obtaining sound recordings of ancient pronunciations. A notable exception is the use of speech recordings in the field of language variation and change,
a branch  of sociolinguistics concerned with small scale variation within communities
(for example, between younger and older members or particular social groups). Some of the
techniques we describe here might also be useful tools to address these kinds of sociolinguistics questions.

Indeed, the analysis of acoustic data highlights one of the
fundamental challenges in comparative linguistics, namely that the
definition of language is an abstraction that simplifies the
reality of speech variability and neglects the somehow continuous geographical spread of spoken varieties, albeit with some clear edges. For example, \citet{Grimes1959} describe as a ``useful
fiction'' the definition of homogeneous speech communities, i.e.
groups of speakers whose linguistic pattern are alike. Given that
for most of human history, most speakers of languages were
illiterate, spoken characteristics are also likely to be of profound
importance in the historical development of languages. The complexity of
the data object (speech) and the large amount of variation call for
careful consideration from the statistical community and we hope
this work will help to arouse attention to this relevant subject.

In the remainder of this paper, we operationalise the term ``language'' to mean a set of recordings of various words in a language or dialect, as spoken on various occasions by a group of speakers, without implying that the vocabulary is complete nor even necessary large. However, the proposed methodology can be applied in a straightforward way to larger and more comprehensive corpora.

We use the expression ``acoustic phonetic data'' to refer to sound
recordings of the same word (or other linguistic unit)  when pronounced by
a group of speakers. In particular, we are interested in the case
where multiple speakers from each language are included in the
data set, since this allows one to better explore statistically the phonetic
characteristics of the language. This is very different from having only
repetitions of a word pronounced by a single speaker and it calls for the development of a novel approach.

The aim of our work is to provide a framework where:
\begin{enumerate}
  \item speech recordings can be analysed to identify features of a
  language,
  \item the variability of speech within the language can be
  considered,
  \item the acoustic differences between languages can be explored on
  the basis of speech recordings, taking into account intra-language
  variability.
\end{enumerate}

Among other things, this will allow us to develop a model (in Section \ref{sec:path}) to explore how the sound
produced by a speaker would be modified when moved towards the
phonetic structure of a different language. More specifically we
will take into account the variability of pronunciation within each
language. This means we explore the variability of the speakers of
the language so that we can then understand where a specific speaker
is positioned in a space of acoustic variation with respect to the population. This allows us
to postulate a path that maps the sound produced by this speaker
to that of a hypothetical speaker with the corresponding position in a
different language. The idea here is to approximate the same kind of
information we can extract when a speaker pronounces words in two
different languages in which they are proficient even if we have
only monolingual speakers. The
observation (audio recordings) of many speakers from each
group is essential to understand the intra-language variability and
thus the relevance of the inter-language acoustic change. This model
has an immediate application in speech synthesis, with the
possibility of mapping a recording from one language to another,
while preserving the speaker's voice characteristics.
This approach could be also extended to modify synthesized speech in such a way that it sounds like the voice of a specific speaker (for example a known actress or a public person). This would be of interest for many commercial applications, from computer gaming to advertising and it is only one
example of the methods that can be developed in the framework we provide. More generally, the framework given here addresses the problem of how to separate speaker-specific voice characteristics from language-specific pronunciation details.

The paper is structured as follows. Section \ref{sec:data} describes
the acoustic phonetic data that are used to demonstrate our methods. We choose to
represent the speech recordings in a time-frequency domain using a
local Fourier transform resulting in surface observations, known in signal processing as spectrograms. Therefore, a short introduction to the functional data
analysis approach to surface data is given in Section \ref{sec:fda}.  The details of these time-frequency representations, as
well as the preprocessing steps needed to remove noise artifacts and
time misalignment between the speech recordings are described in
Section \ref{sec:preproc}. Section \ref{sec:est} illustrates how to
estimate some crucial functional parameters of the population of log-spectrograms and claims that the covariance structures are common
across all the words in each language. Section \ref{sec:path} is
devoted to the definition and exploration of cross-linguistic
phonetic differences and shows how the pronunciation of a word can be morphed into another language while preserving the speaker/voice identity. The final section gives a discussion of the
advantages of the proposed method and of how it is possible to extend
it to even more complex situations, where the phonetic features
depend continuously on historical or geographical variables.

\section{The Romance digits data set}
\label{sec:data} The methods in this paper will be illustrated with
an application to a data set of audio recordings of digits in Romance languages. This data set
was compiled in the Phonetics Laboratory of the University of
Oxford in 2012-2013. It consists of natural speech recordings
of five languages: French, Italian, Portuguese, American Spanish and
Castilian Spanish, the two varieties of Spanish being considered
different languages for the purpose of the analysis. The speakers
utter the numbers from one to ten in their native language. The data set
is inherently unbalanced; we have seven French speakers, five
Italian speakers, five American Spanish speakers, five Iberian
Spanish speakers and three Portuguese speakers,
resulting in a sample of $219$ recordings. The sources of the
recordings were either collected from freely available language training websites or standardized recordings made by
university students. As this data set consists of recordings made
under non-laboratory settings, large variabilities may be
expected within each group. This provides a real-world setting for our analysis, and allows us to build models which characterise realistic variation in speech recording, somewhat of a prerequisite for using this model in practice, as field work recordings are often not recorded under laboratory conditions. The data set is also heterogeneous in terms of sampling rate, duration and format. As such, before any
phonetic or statistical analysis took place, all data were converted
to $16$-bit PCM (pulse code modulation) *.wav files at a sampling rate of $16$ kHz. We indicate each sound recording as
$x^L_{ik}(t)$, where $L$ is the language, $i=1,\ldots,10$ the
pronounced word and $k=1,\ldots,n_L$ the speaker, $n_L$ being the
number of speakers available for language $L$, and $t$ time. This
data set has been collected within the scope of \textit{Ancient Sounds}, a
research project with the aim of regenerating audible
spoken forms of the (now extinct) earlier versions of Indo-European
words, using contemporary audio recordings from multiple languages.
More information about this project can be found on the website
\url{http://www.phon.ox.ac.uk/ancient_sounds}.

Although the cross-linguistic comparison of spoken digits is
interesting in its own right, this subset of words can also be
considered as representative of a language's vocabulary from a
phonetic point of view, meaning that the words used for the numbers
in the Romance languages were not chosen to possess any specific
phonetic structure. Consequently, we use the word ``language'' as a
shorthand for these particular small samples of digit
recordings. However, we view this analysis as a proof of concept,
and will not focus on the problem of the representativeness of the
sample of speakers or words. In view of a broad possible application
of the approach which will be outlined, more structured choices of
representative words could be taken or specific dialect choices
made, but the approach would remain the same.

\section{The analysis of surface data}
\label{sec:fda}

Different representations are available in phonetics to deal
with speech recordings \citep[see, e.g.][]{cooke1993}. Many of them share the idea of
representing the sound with the distribution of intensities over
frequency $\omega$ and time $t$.  We choose in particular the power spectral
density of the Local Fourier Transform (i.e. the narrow-band Fourier spectrogram), as detailed in Section
\ref{sec:preproc}. This widely-used representation is a
two-dimensional surface that describes the sound intensity for each
time sample in each frequency band. Since we can represent each spoken word
as a two dimensional smooth surface, it is natural to employ a
functional data analysis approach. Good results have already been
obtained applying functional data analysis techniques to acoustic
analysis, although in the different context of single language
studies, for example in \citet{Koenig2008} and
\citet{Hadjpantelis2012}. Functional data analysis is appropriate in
this context because it addresses problems where data are
observations from continuous underlying processes, such as
functions, curves or surfaces. A
general introduction to the analysis of functional data can be found
in \citet{Ramsay2005} and in \citet{Ferraty_Vieu_2006}. The central
idea is that taking into account the smooth structure of the process
helps in dealing with the high dimensionality of the data objects.

In contrast, in most previous quantitative work on pronunciation variation, such as sociolinguistics or experimental phonetics, only one or a few acoustic parameters (one-dimensional time series) are examined, e.g. pitch or individual resonant frequencies. Variations in vowel qualities, for example, are typically represented by just two data points: the lowest two resonant frequencies (the first and second formants) measured at the mid-point of the vowel. Such a two-dimensional representation lends itself well to simple visualization of a large number of observations, in the form of a scatterplot. But although the validity of two-frequency representations of vowels or single-variable representations of pitch or loudness is motivated by decades of prior research, it clearly suffers from two limitations. First, almost all of the available time-frequency-amplitude information in the speech signal is simply discarded as if it were irrelevant. Second, we do not always know in advance which acoustic parameters are most relevant to a particular investigation; therefore, a more holistic approach to analysis of speech signals may be helpful. The methods presented in this paper, which take the entire spectrogram of each audio recording as data objects, enable us to examine and to manipulate a variety of properties of speech that are not easily reduced to a single low-dimensional data point. By considering higher-order statistical properties of the shape of spectrograms, it becomes possible to characterise such notions as the typical pronunciation of a word, what each speaker sounds like (in general, irrespective of what words they are saying), how their pronunciation differs from that of other speakers, and what it is that makes two languages sound different, above and beyond the differences in the words they use and the speakers involved.

More formally, we consider here data objects that are two dimensional surfaces on
a bounded domain, as it is the case of spectrograms. Let $X$
be a random surface so that $X\in L^2([0,\Omega]\times
[0,T])$ and $E[||X^2||^2]<+\infty$. A mean surface can then be
defined as $\mu(\omega,t)=E[X(\omega,t)]$ and the four dimensional
covariance function as
$c(\omega,\omega',t,t')=\mathrm{cov}[X(\omega,t),X(\omega',t')]$.

In practice these surfaces are observed over a finite number of grid
points and they are affected by noise; indeed they can be thought of as a noisy image. As noted by
\citet{Ramsay2005}, ``the term \emph{functional} in reference to
observed data refers to the intrinsic structure of the data rather
than to their explicit form''. Thus a smoothing step is needed to
recover the regular surfaces that reflect the properties of the
underlying process. These surfaces are represented by means of a
linear combination of basis functions which span the separable
Hilbert space $L^2([0,\Omega]\times [0,T])$. In particular, we choose the
widely popular method of smoothing splines to estimate a smooth
surface $\widetilde{X}(\omega,t)$ from the noisy observation on a
regular grid $\mathfrak{X}(\omega_i,t_j),\ i=1,\ldots,n_{\omega},\
j=1,\ldots,n_t$.

When analysing a sample of surfaces, we are implicitly assuming that
the comparison of their values at the same coordinates $(\omega,t)$
is meaningful. However, this is often not the case when data are
measurements of a continuous process such as human speech. For
example, different speakers (or even the same speaker in different
replicates) can speak faster or slower without this changing the meaningful
acoustic information in the recordings. The resulting sound objects
are obviously not comparable though, unless this problem is
addressed first. This situation is so common in functional data
analysis that much work has been devoted to its solution and these
techniques are referred to as functional registration (or warping or
alignment, see \citet{Marron2014} and references therein for details). In the case of a two dimensional surface, the
misalignment can in principle affect both coordinates; this is the
case for example in image processing. A two dimensional warping
function $h(\omega,t)$ is then needed to align each surface and this
is a more complex problem than one-dimensional registration.
However, even though we are considering data that are surfaces, the
way they are produced, which will be detailed in Section
\ref{sec:preproc}, makes it sensible to adjust only for the misalignment on the temporal axis, this being due to different speaking rates, which are not relevant for our goals. On the contrary, we want to preserve the differences on the frequency axis which contains information about the phonetic characteristics of the speakers.

Thus, we apply a mono-dimensional warping to our surface data.
If we aim to align a sample of surfaces
$\widetilde{X}_1,\ldots,\widetilde{X}_N$, we look for a set of
time-warping functions $h_1(t),\ldots,h_N(t)$ so that the aligned
surface will be defined as $X_1=\widetilde{X}_1(\omega,h_1(t)),\ldots,
X_N=\widetilde{X}_N(\omega,h_N(t))$. In the next section we will
describe how to achieve this in practice for acoustic phonetic data.

Given the smooth and aligned surfaces $X_1,\ldots, X_N$, it is
possible to estimate the functional parameters of the underlying
process, for example
$$
\widehat{\mu}(\omega,t)=\frac{1}{N}\sum_{i=1}^N X_i,\ \ \
\widehat{c}(\omega,\omega',t,t')=\frac{1}{N-1}\sum_{i=1}^N
(X_i(\omega,t)-\widehat{\mu}(\omega,t))(X_i(\omega',t')-\widehat{\mu}(\omega',t')).
$$
However, the high-dimensionality of the problem makes the estimate
for the covariance structure inaccurate or even computationally
unfeasible. In Section \ref{sec:est} we introduce some modelling
assumptions to make the estimation problem tractable.

\section{From speech records to smooth spectrogram surfaces}
\label{sec:preproc}

As mentioned in the previous section, we choose to represent the
sound signal via the power spectral density of the local Fourier
transform.  This means we first apply a local Fourier transform to
obtain a two dimensional spectrogram which is function of time (the
time instant where we centre the window for the local Fourier
transform) and frequency. For the Romance digit data, we
use a Gaussian window function $w$ with a window length of ten
milliseconds (a reasonable length for the signal to be considered as stationary), defined as
$\psi(\tau)=\exp(-\frac{1}{2}(\frac{\tau}{0.005})^2)$. Since the
original acoustic data  was sampled at $16$ kHz, this results in a
window size of 160 samples per frame and the maximal effective
frequency detected is $8$ kHz, the Nyquist frequency of our sampling
procedure.

\begin{figure}[h!]
\includegraphics[height=0.3\textwidth,width=0.7\textwidth]{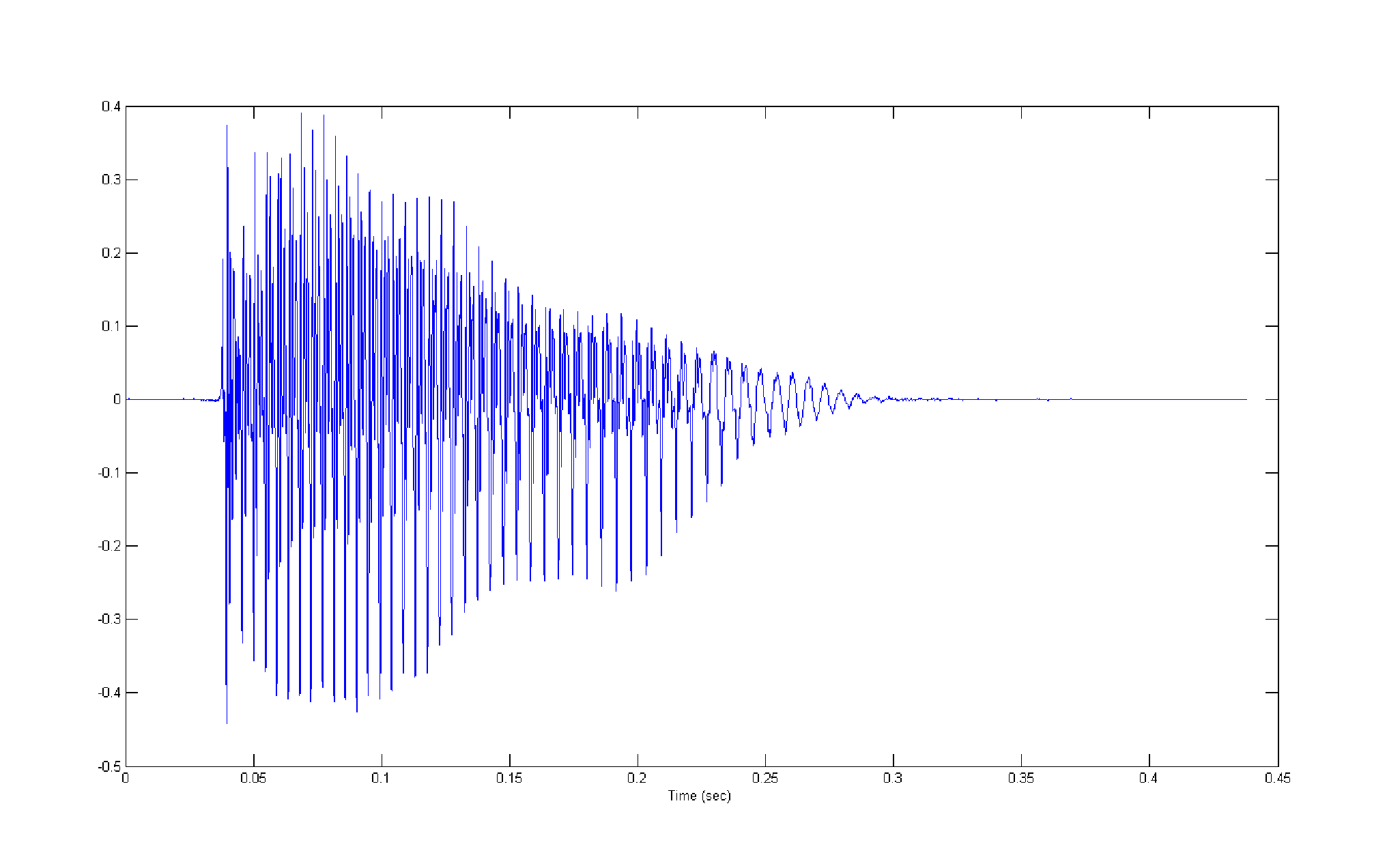}
\centering
\includegraphics[height=0.28\textwidth,width=0.35\textwidth]{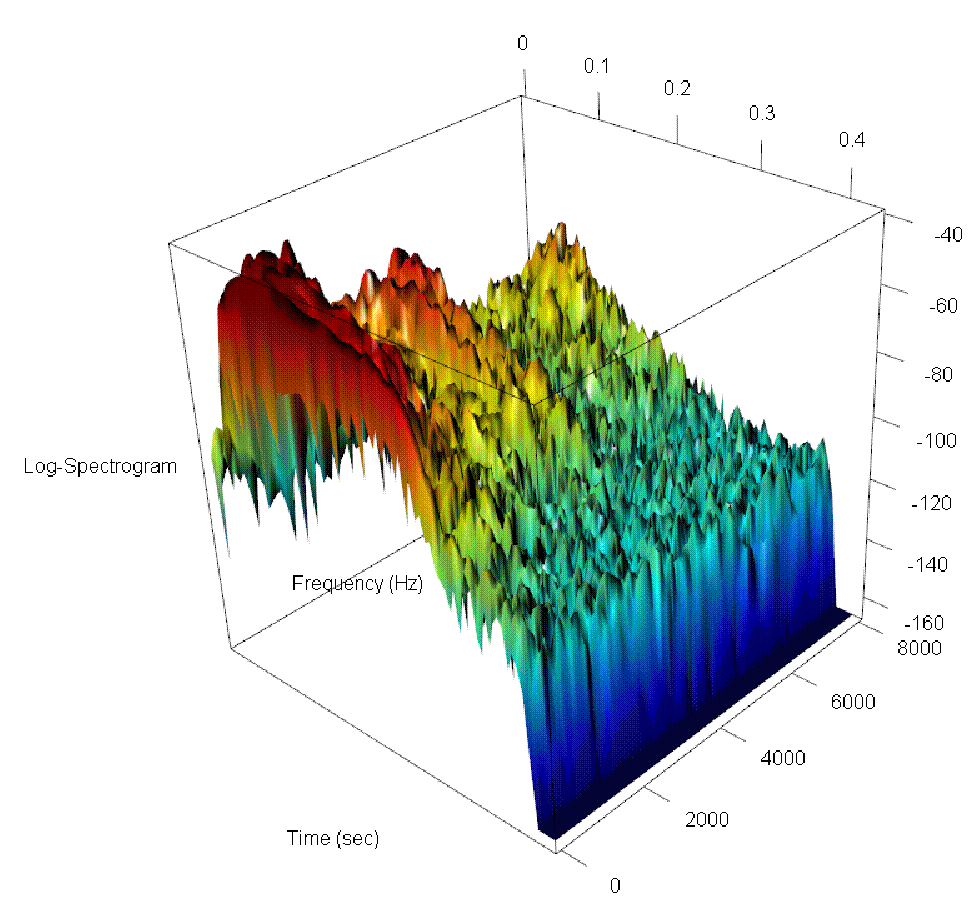}
\includegraphics[height=0.3\textwidth,width=0.4\textwidth]{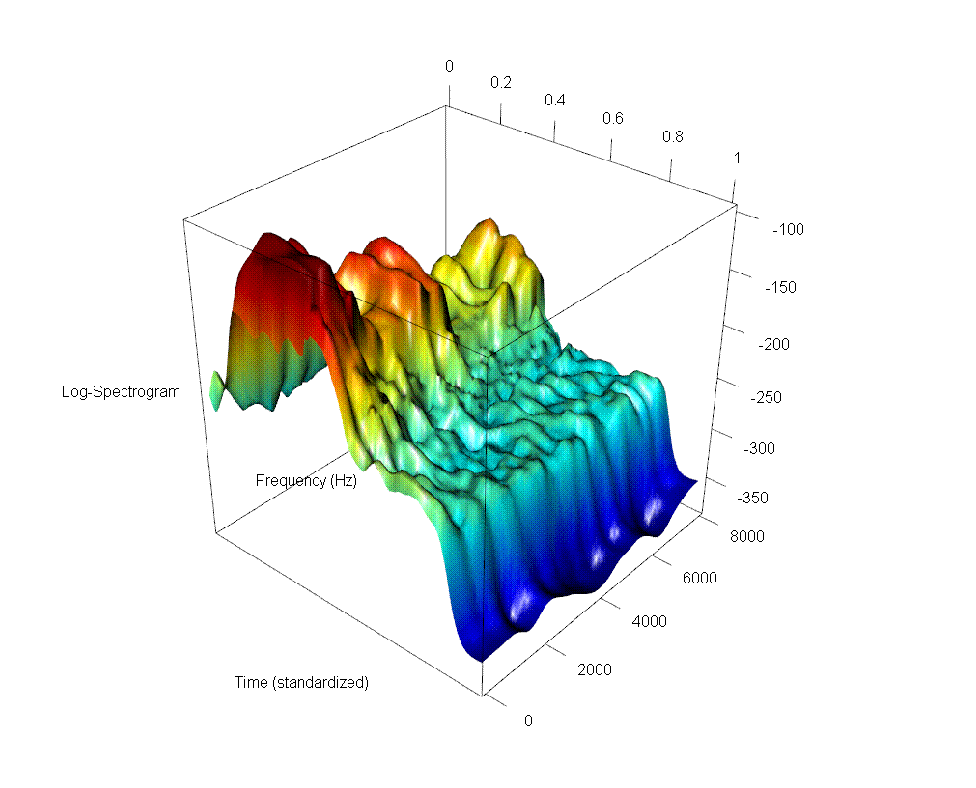}
\caption{Raw record (top), raw log-spectrogram (bottom left) and
smoothed and aligned log-spectrogram (bottom right) for a French
speaker pronouncing the word ``un'' (``one'').} \label{fig:spectro}
\end{figure}

We can compute the local Fourier transform at angular frequency $\omega$ and time $t$ as
$$
X^L_{ik}(\omega,t)=\int_{-\infty}^{+\infty}
x^L_{ik}(\tau)\psi(\tau-t)e^{-j\omega\tau}\mathrm{d}\tau.
$$
The power spectral density, or spectrogram, defined as the
magnitude of the Fourier transform and the log-spectrogram (in
decibel), is therefore
$$
\mathfrak{S}^L_{ik}(\omega,t)=10\log_{10}(|X^L_{ik}(\omega,t)|^2).
$$
Fig.\ \ref{fig:spectro} shows an example of a raw speech signal
(top panel) and the corresponding log-spectrogram (bottom left
panel), for the sound produced by a French speaker pronouncing the
word \textit{un} [\~\oe],

To deal with these objects in a functional way, we need to address the
problems of smoothing and registration described in the previous
section. Indeed, when data comes from real world recordings, as
opposed to laboratory conditions, the raw log-spectrograms suffer
from noise. For this reason we apply a penalized least
square filtering for grid data using discretized smoothing splines.
In particular, we use the automated robust algorithm for two-dimensional gridded data described in
\citet{Garcia2010}, based on the discrete cosine transform, which
allows for a fast computation in high dimensions when the grid is
equally spaced. The smoothed log-spectrograms can be transformed back into sounds (via Inverse Fourier Transform) to check that the smoothing removed the noise without loss of any relevant linguistics information.

The second preprocessing step consists of registration in time. This is necessary because speakers can speak faster or slower and this is
particularly true when data are collected from different sources
where the context is different. However, differences in the
speech rate are normally not relevant from a linguistic point of view and
thus alignment along the time axis is needed because of this
time misalignment in the acoustic signals. First, we standardized the
time scale so that each signal goes from $0$ to $1$. Then, we adapt
to the case of surface data the procedure proposed in
\citet{TangM2008} to remove time misalignment from functional
observations. Given a sample of functional data $f_1,\ldots,f_n\in
L^2([0,1])$, this procedure looks for a set of strictly monotone time
warping functions $h_1,\ldots,h_n$ so that $h_i(0)=0$, $h_i(1)=1$,
$i=1,\ldots,n$. In practice, these warping functions are modelled via
spline functions and estimated by minimizing the pairwise
difference between the observed curve while penalizing their
departure from the identity warping $h(t)=t$. Hence, a pairwise
warping function is first obtained as
$$
h_{ij}(t)=\arg \min_{h} \int_{0}^1 (f_i(h(t))-f_j(t))^2 + \lambda
\int_0^1 (h(t)-t)^2,
$$
where the minimum is computed over all the spline functions
on a chosen grid. Now let $h_k$, $k=1,\ldots,n$, be the warping
function from a specific time to the standardized time
scale. Then, if $s=h_j^{-1}(t)$,
$h_i(s)=h_i(h_j^{-1}(t))=h_{ij}(t)$. Under the assumption of the
warping function to have the identity on average and thus
$E[h_{ij}|h_j]=h_j^{-1}$, the estimator proposed by
\citet{TangM2008} is
$$
h_j^{-1}(t)=\frac{1}{n}\sum_{j=1}^n h_{ij}(t).
$$

To apply this idea to acoustic phonetic
data, we need first to define the groups of log-spectrograms we want
to align together. As the mean log-spectrogram is
different from word to word, we decide to align the log-spectrograms
corresponding to the same word. Then, we have to extend the
procedure to two dimensional objects such as surfaces. As mentioned
in the previous section, it is safe to assume that there is no phase
distortion in the frequency direction, given the
relatively narrow window used in the local Fourier transform. In contrast, time misalignment can be a serious issue due to
differences in speech rate across speakers. Therefore we modify the
procedure in \citet{TangM2008} so that we look for pairwise time
warping functions but minimize the discrepancy between surfaces.
For each word $i=1$ in a group of log-spectrograms we want to align, for every pair of languages $L$ and $L^\prime$ and for every pair of speakers $k$ and $m$,
we define the discrepancy between the log-spectrogram
$\widetilde{\mathfrak{S}}^L_{ik}$ and
$\widetilde{\mathfrak{S}}^{L^\prime}_{im}$ as
\begin{equation}
\label{eq:discrepancy}
D_{\lambda}(\widetilde{\mathfrak{S}}^L_{ik},\widetilde{\mathfrak{S}}^{L^\prime}_{im},g^{LL^\prime}_{km})
=    \int_{f=0}^{+\infty}
\int_{t=0}^{1}(\widetilde{\mathfrak{S}}^L_{ik}(\omega,
g^{LL\prime}_{km}(t))-\widetilde{\mathfrak{S}}^{L^\prime}_{im}(\omega,
t))^2 +  \lambda(g^{LL^\prime}_{km}(t) - t)^2  dt d \omega,
\end{equation}
where $\lambda$ is an empirically evaluated non-negative
regularization constant and $g^{LL^\prime}_{km}(\cdot)$ is the
pairwise warping function mapping the time evolution of
$\widetilde{\mathfrak{S}}^L_{ik}(\omega, t)$ to that of
$\widetilde{\mathfrak{S}}^{L^\prime}_{im}(\omega, t)$. We obtain the
pairwise  warping function $\widehat{g}^{LL^\prime}_{km}(\cdot)$ by
minimizing the discrepancy
$D_{\lambda}(\widetilde{\mathfrak{S}}^L_{ik},\widetilde{\mathfrak{S}}^{L^\prime}_{im},g^{LL^\prime}_{km})$
under the constraint that $g^{LL\prime}_{km}$ is piecewise-linear,
monotonic and so that $g^{LL^\prime}_{km}(0)=0$ and
$g^{LL^\prime}_{km}(1)=1$. Finally, the inverse of the global
warping function for each pronounced word can be estimated as the
average of the pairwise warping functions:
$$
h^{-1}_{ik}=\frac{1}{\sum_{L^\prime=1}^5 n_L}\sum_{L^\prime=1}^5
\sum_{m=1}^{n_L} \widehat{g}^{LL^\prime}_{km},
$$
and the smoothed and aligned log-spectrogram for the language $L$,
word $i$ and speaker $k$ is therefore $ S_{ik}^L (\omega,
t)=\widetilde{\mathfrak{S}}^L_{ik}(\omega, h_{ik}(t))$. In practice, warping functions are represented with a spline basis defined over a regular grid of $100$ points on $[0,1]$ and we look for the spline coefficients that minimize the discrepancies. The quantities in (\ref{eq:discrepancy}) are approximated by their discretized equivalent on a two-dimensional grid with $100$ equispaced grid points on the time dimension and $81$ equispaced grid points in the frequency dimension. In general, the number of grid points in the time axis needs to be chosen based on the length of the uttered sounds but we have seen that $100$ points provide an accurate  reconstruction of the log-spectrograms in the Romance digit dataset.

After this second preprocessing step, we are presented with $219$
smoothed and aligned log-spectrograms. For example, the smoothed and
time-aligned log-spectrogram from the sound produced by a French
speaker pronouncing the word \textit{un} can be found in the bottom
right panel of Fig.\ \ref{fig:spectro}.

Other choices are of course possible in the preprocessing of the speech data. In particular, the time registration based on the minimization of the Fisher-Rao metric \citep{Srivastava2011} can be a computationally more efficient alternative when computing time is of concern. By way of example, we also replicate the analysis of the Romance digit data when the smoothing is performed with the thin-plate regression splines implemented in the R package \texttt{mgcv} \citep{Wood2003} and the time registration is obtained by minimising the Fisher-Rao metric \citep[R package \texttt{fdasrvf},][]{fdasrvf}. The results of this alternative analysis are available as supplementary material and are qualitatively similar to the one reported below, giving credence to the idea that the results are not simply systematic mis-registration by one technique versus another.

\section{Estimation of means and covariance operators}
\label{sec:est}

The process that generates the sounds (and thus their representation
as log-spectrograms) is governed by unknown parameters that depend
on the language, the word being pronounced and the speaker. However,
we need to make some assumptions to identify and estimate these
parameters. We consider the mean log-spectrogram as
depending on the particular word in each language being pronounced. Indeed,
the mean spectrogram is in general different for the different
words, as would be expected. Let
$i=1,\ldots,10$ be the pronounced words and $k=1,\ldots,n_L$ the
speakers for the language $L$. The smoothed and aligned
log-spectrograms $S_{ik}^L(\omega,t)$ allow the estimation of the
mean log-spectrogram
$\overline{S}_{i}^L(\omega,t)=(1/n_{L})\sum_{k=1}^{n_L}
S_{ik}^L(\omega,t)$ for each word $i$ of the language $L$.

Recent studies \citep[][]{Aston2010,Pigoli2014} show that
significant linguistic features can be found in the covariance
structure between the intensities at different frequencies. This can
be considered as a summary of what a language ``sounds like'', without
incorporating the differences at the word level. Thus, we first
assume in our analysis that the covariance structure of the
log-spectrograms is common for all the words in the language and we estimate it using the residual surface obtained by
removing the word mean effect. In Section \ref{sec:test} we develop
a procedure to verify this assumption in the Romance digit
data set.

Starting from the smoothed and aligned log-spectrograms
$S_{ik}^L(\omega,t)$ of the records of the number $i=1,\ldots,10$
for the speakers $k=1,\ldots,n_i$, we thus focus on the residual
log-spectrograms $R_{ik}^L(\omega,t)=S_{ik}^L(\omega,t)-
(1/n_{i})\sum_{k=1}^{n_i} S_{ik}^L(\omega,t)$, which measure how
each token differs from the word mean. In the following, we
disregard in the notation the different speakers and words and for the residual log-spectrogram indicate by
$R_{j}^L(\omega,t), j=1,\ldots,n_L$ the set of observations for the
language $L$ including all speakers and words.

However, using standard covariance estimation techniques to find the full four-dimensional covariance
structure is  computationally expensive or not statistically feasible (because of the small sample size), thus we need some
modelling assumptions. There are many ways to incorporate assumptions which allow such estimation, a common one being some form of sparsity. Rather than the usual definition of sparsity that many elements are zero, we prefer to work on the principle that the covariance can be factorised.

We assume that the covariance structure
$c^L(\omega_1,\omega_2,t_1,t_2)=cov(S^L(\omega_1,t_1),S^L(\omega_2,t_2))$
is separable in time and frequency, i.e.
$c^L(\omega_1,\omega_2,t_1,t_2)=c^L_{\omega}(\omega_1,\omega_2)c^L_t(t_1,t_2)$.
While we do not necessarily believe this assumption to be true in
general, a structure is needed to obtain reliable estimates for
the covariance operators, and it is a reasonable assumption that is frequently (implicitly) used in signal processing, particularly when constructing higher dimensional bases from lower dimensional ones.

Possible estimates for $c^L_{\omega}(\omega_1,\omega_2)$ and
$c^L_{t}(t_1,t_2)$  are
\begin{equation}
\label{eq:sep_est}
\widehat{c}^L_{r}=\frac{\widetilde{c}^L_{r}}{\sqrt{\mathrm{trace}(\widetilde{c}^L_{r})}},\hspace{0.1cm}
r=\omega,t,
\end{equation}
where $\mathrm{trace}$ indicates the trace of the covariance
function, defined as $\mathrm{trace}(c)=\int c(s,s)\mathrm{d}s$,
while $\widetilde{c}^L_{r}$, $r=\omega,t$ are the sample marginal
covariance functions
$$
\widetilde{c}^L_{\omega}(\omega_1,\omega_2)=\frac{1}{n_L-1}\sum_{j=1}^{n_L}
\int_{0}^{1}
(R^L_{j}(\omega_1,t)-\overline{R}^L_{n_L}(\omega_1,t))(R^L_{j}(\omega_2,t)-\overline{R}^L_{n_L}(\omega_2,t))\mathrm{d}t,
$$
and
$$
\widetilde{c}^L_{t}(t_1,t_2)=\frac{1}{n_L-1}\sum_{j=1}^{n_L}
\int_{0}^{8\mathrm{kHz}}
(R^L_{j}(\omega,t_1)-\overline{R}^L_{n_L}(\omega,t_1))(R^L_{j}(\omega,t_2)-\overline{R}^L_{n_L}(\omega,t_2))\mathrm{d}\omega,
$$
$\overline{R}^L_{n_L}$ being the sample mean of the residual
log-spectrogram for the language $L$. We introduce also the
associated covariance operators as
$$
\widehat{C}^L_r
g(x)=\int_{0}^{M}\widehat{c}^L_{r}(x,x')g(x')\mathrm{d}x'
\hspace{4mm} g \in L^2(\mathbb{R}),\hspace{0.1cm} (r,M)\in\{(\omega,8\mathrm{kHz}),(t,1)\}.
$$

To see why we choose (\ref{eq:sep_est}) to estimate the
two separable covariance functions, let $\widetilde{c}^L_{\omega}$
and $\widetilde{c}^L_{t}$ be the true marginal covariance functions,
i.e.
$$
\widetilde{c}^L_{\omega}(\omega_1,\omega_2)=\int_{0}^{1}
c^L(\omega_1,\omega_2,t,t)\mathrm{d}t,\ \ \
\widetilde{c}^L_{t}(t_1,t_2)=\int_{0}^{8\mathrm{kHz}}
c^L(\omega,\omega,t_1,t_2)\mathrm{d}\omega.
$$
Then, if the full covariance function is indeed separable, their
product can be rewritten as
$$
\widetilde{c}^L_{\omega}(\omega_1,\omega_2)
\widetilde{c}^L_{t}(t_1,t_2)=\int_{0}^1
c^L_{\omega}(\omega_1,\omega_2)c^L_t(t,t)\mathrm{d}t
\int_{0}^{8\mathrm{kHz}}c^L_{\omega}(\omega,\omega)c^L_t(t_1,t_2)\mathrm{d}\omega=$$
$$
=c^L_{\omega}(\omega_1,\omega_2)\mathrm{trace}(c^L_t)c^L_t(t_1,t_2)\mathrm{trace}(c^L_{\omega}).
$$
Moreover,
$\mathrm{trace}(\widetilde{c}^L_{\omega})=\mathrm{trace}(c^L_{\omega}\mathrm{trace}(c^L_t))=\mathrm{trace}(c^L_{\omega})\mathrm{trace}(c^L_t)$
and the same is true for $\widetilde{c}^L_t$. Hence,
$$
\frac{\widetilde{c}^L_{\omega}(\omega_1,\omega_2)}{\sqrt{\mathrm{trace}(\widetilde{c}^L_{\omega})}}\frac{\widetilde{c}^L_t(t_1,t_2)}{\sqrt{\mathrm{trace}(\widetilde{c}^L_t)}}=c^L_{\omega}
(\omega_1,\omega_2) c_t^L(t_1,t_2)=c^L(\omega_1,\omega_2,t_1,t_2)
$$
and this suggests $\widehat{c}^L_{r}$ as estimator for $c^L_r$,
$r=\omega,t$.

\begin{figure}[h]
\centering
\includegraphics[height=0.3\textwidth,width=0.3\textwidth]{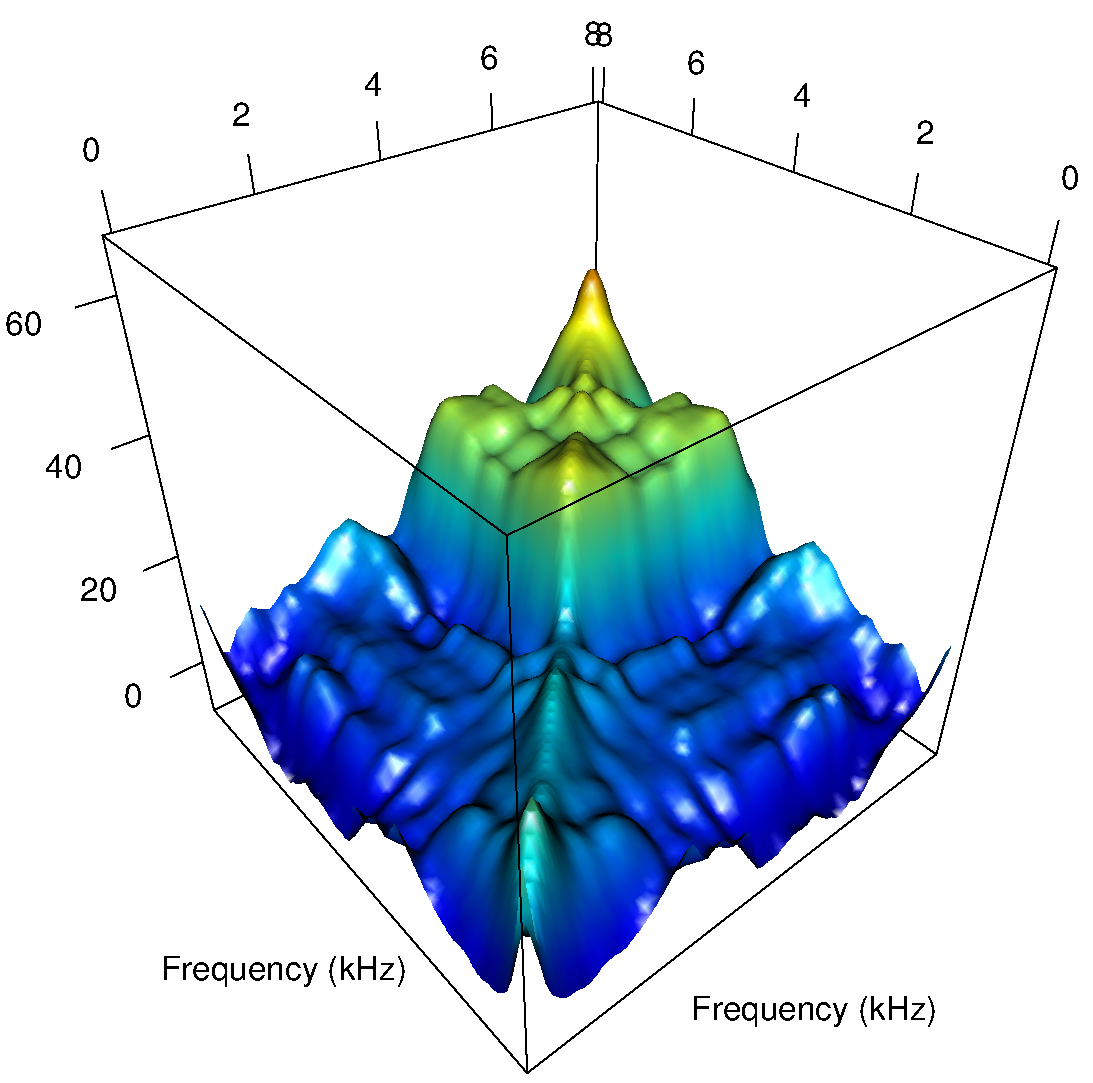}
\includegraphics[height=0.3\textwidth,width=0.3\textwidth]{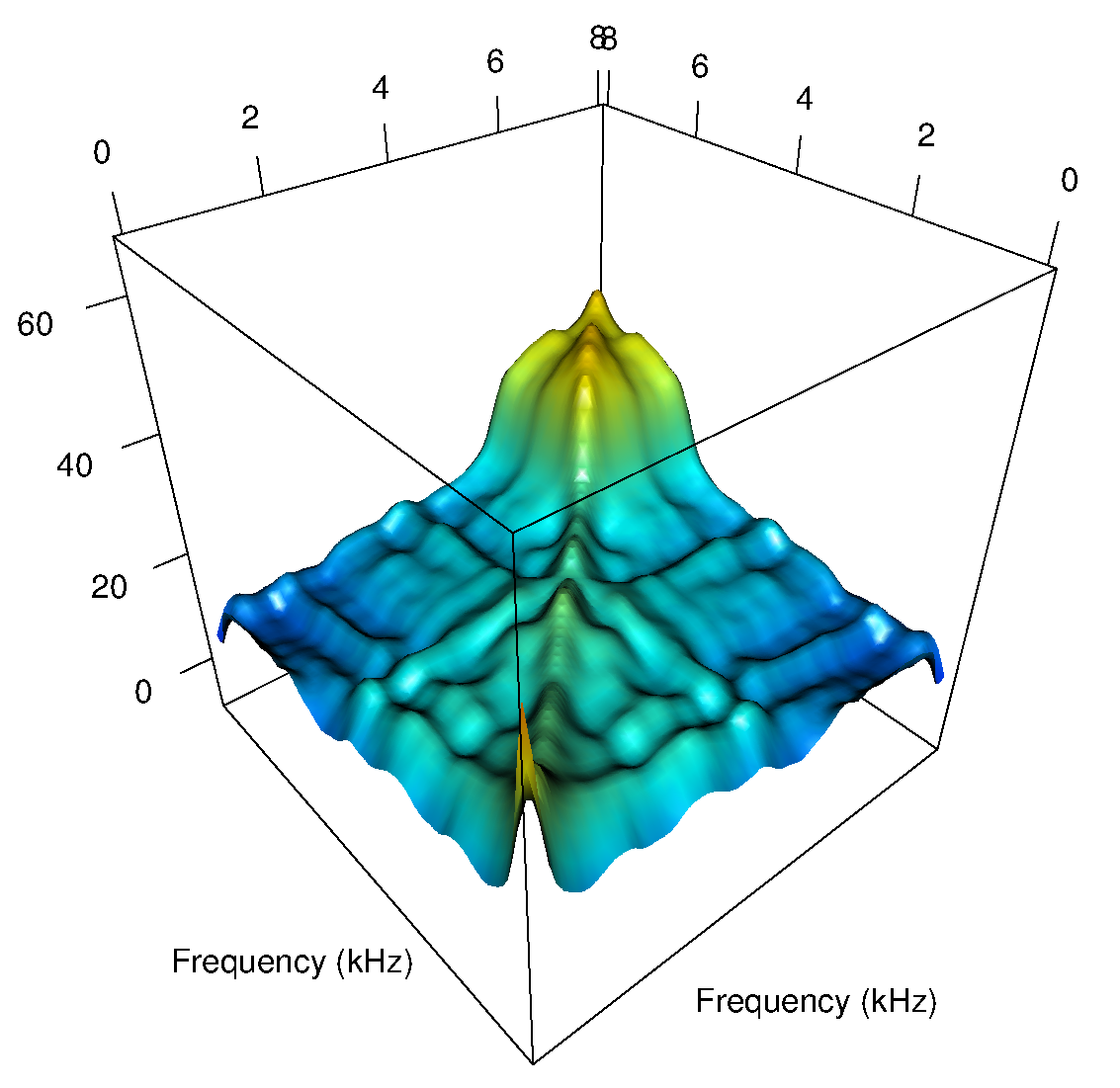}
\includegraphics[height=0.3\textwidth,width=0.3\textwidth]{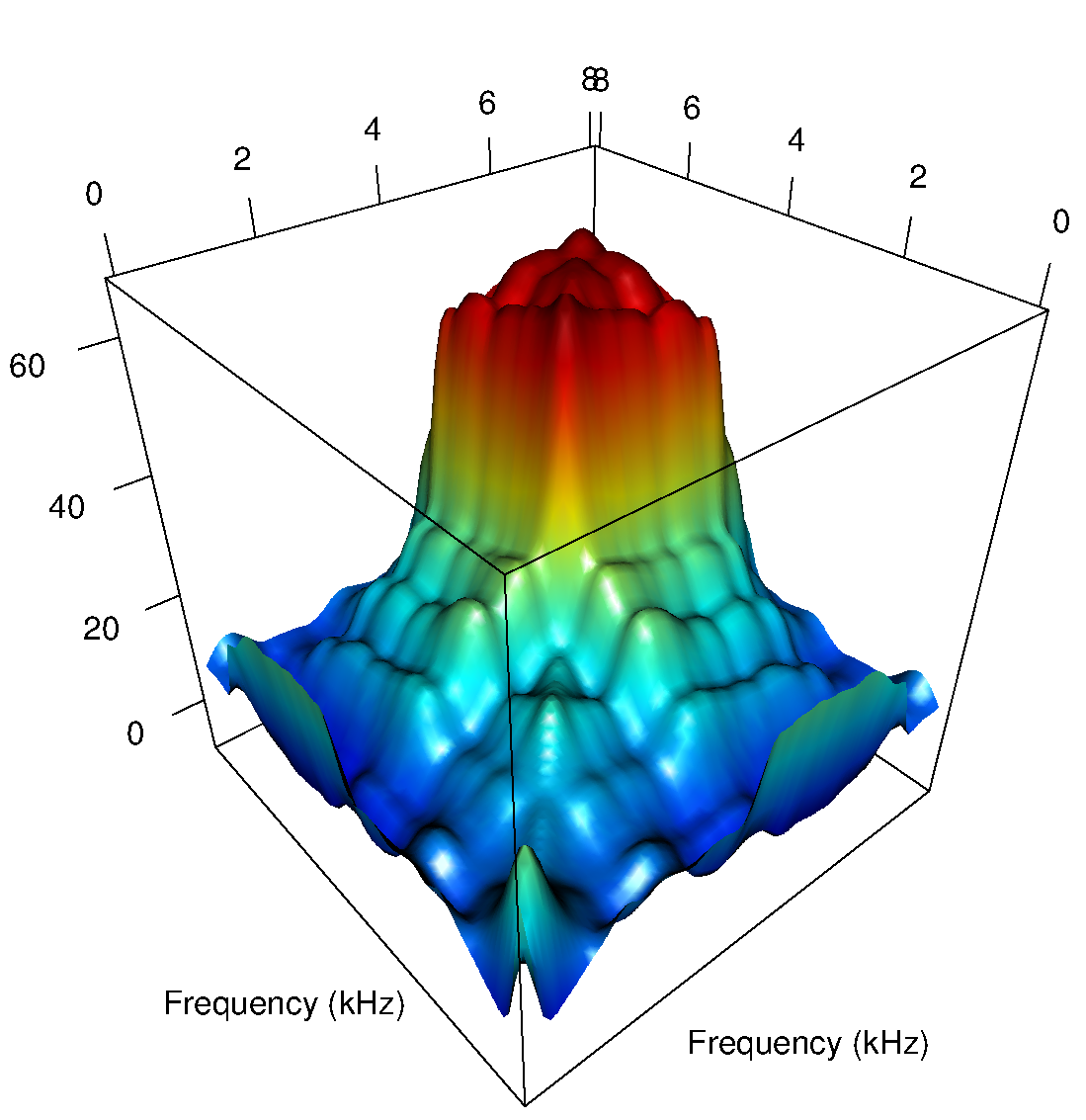}
\includegraphics[height=0.3\textwidth,width=0.3\textwidth]{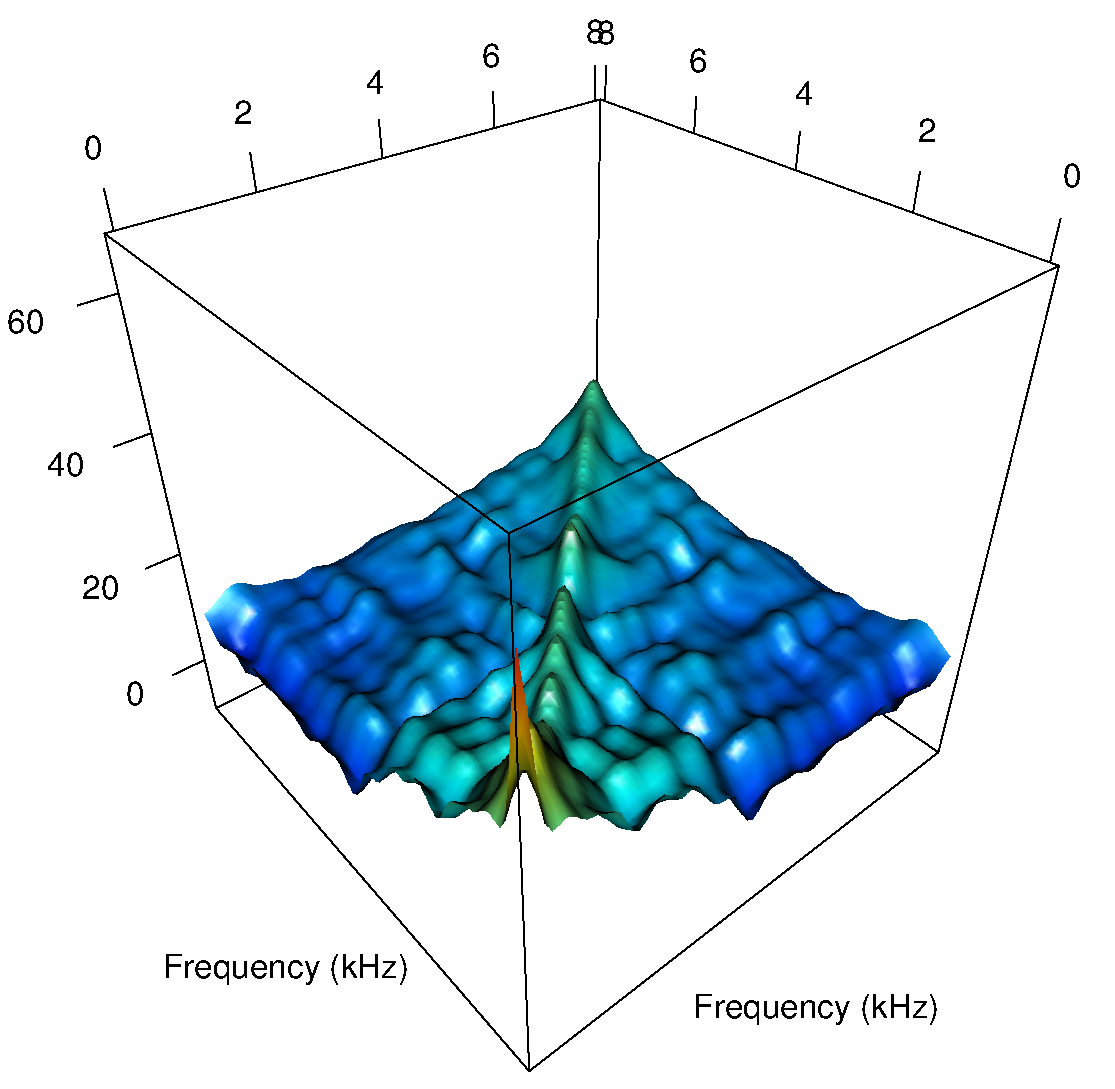}
\includegraphics[height=0.3\textwidth,width=0.3\textwidth]{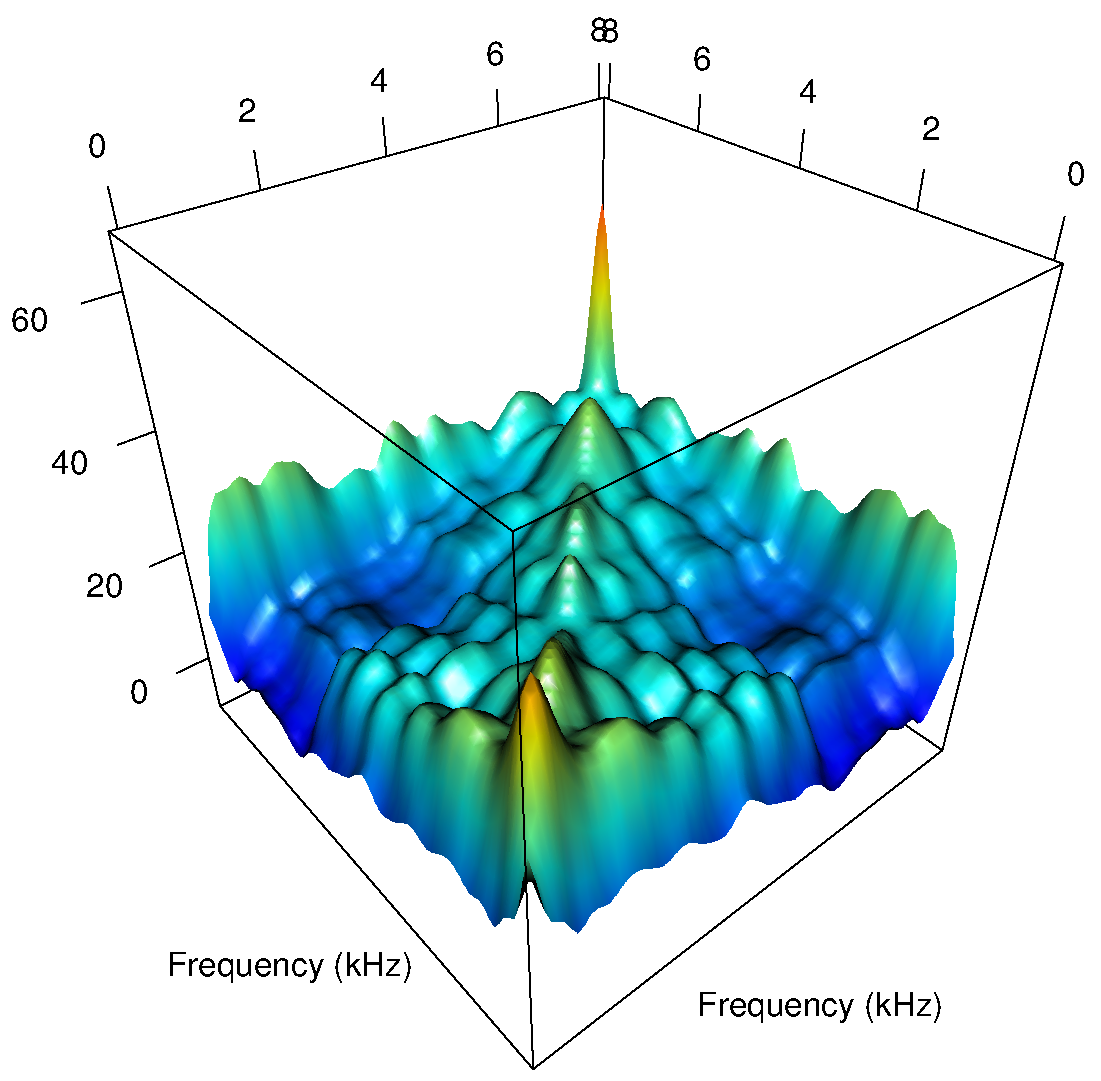}
\caption{Marginal covariance function between frequencies for the
five Romance languages. First row: Italian (left), French (center)
and Portuguese (right). Second row: American Spanish (left) and
Iberian Spanish (right).} \label{fig:lang}
\end{figure}

\begin{figure}[h]
\centering
\includegraphics[height=0.3\textwidth,width=0.3\textwidth]{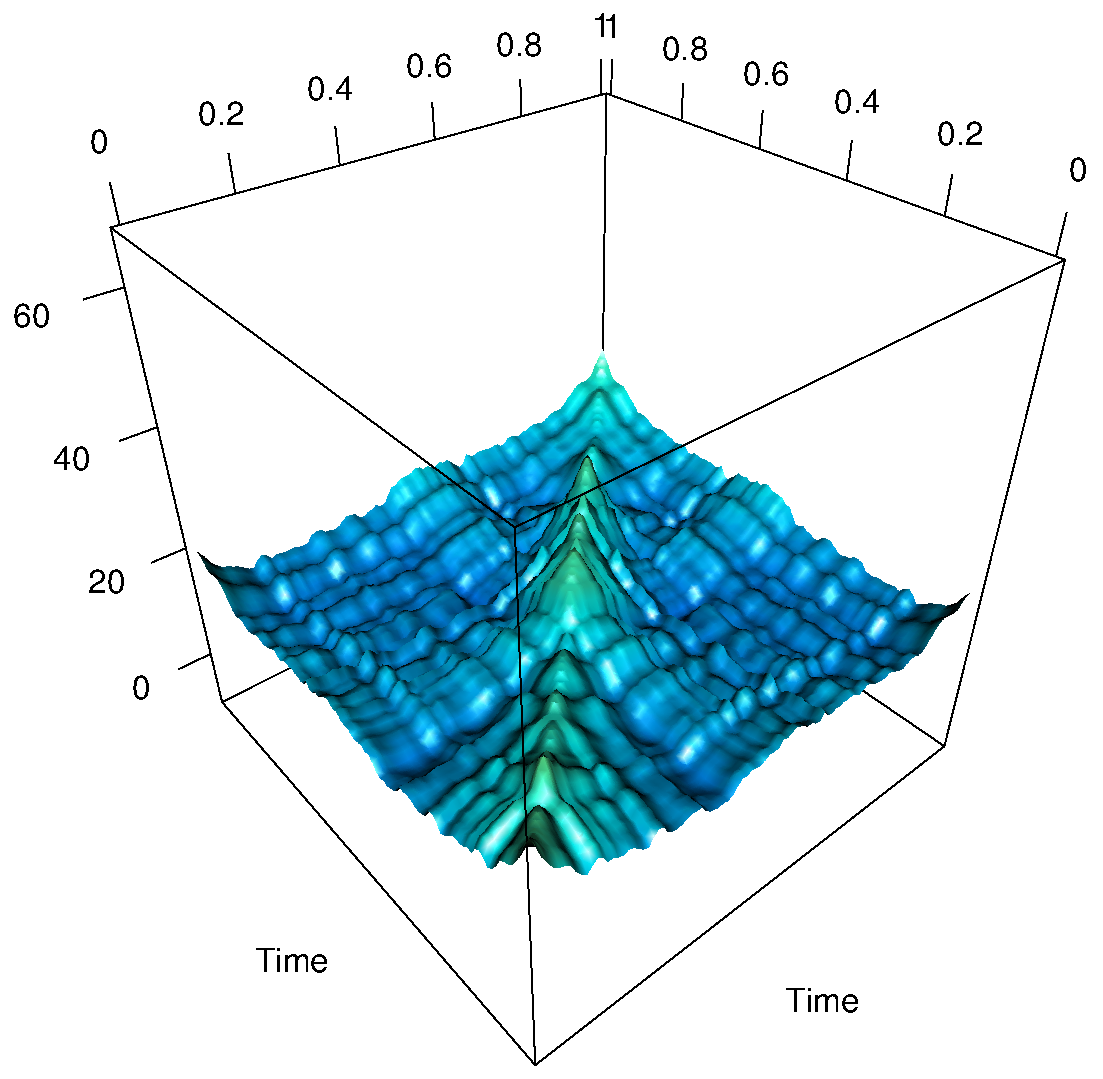}
\includegraphics[height=0.3\textwidth,width=0.3\textwidth]{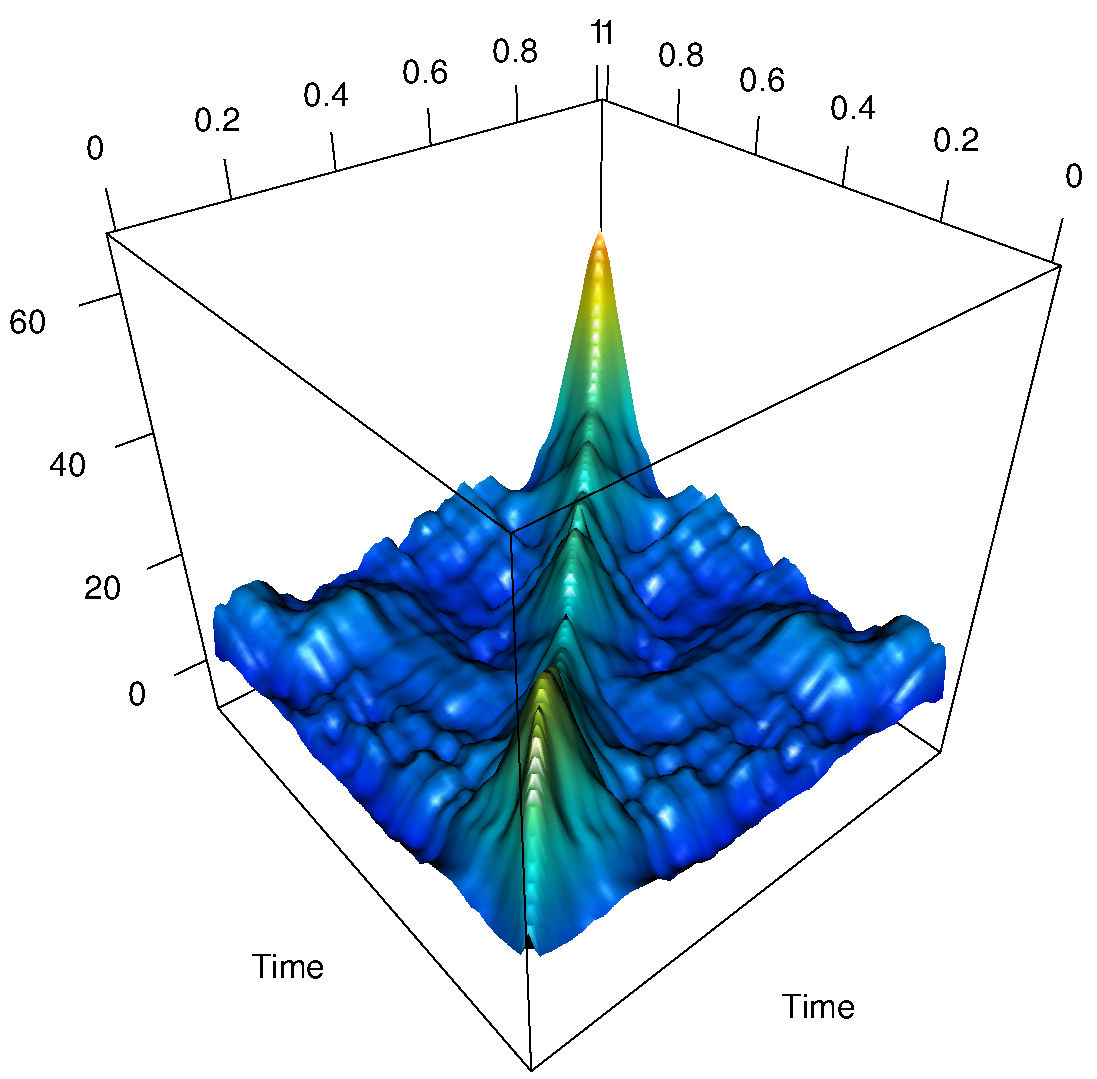}
\includegraphics[height=0.3\textwidth,width=0.3\textwidth]{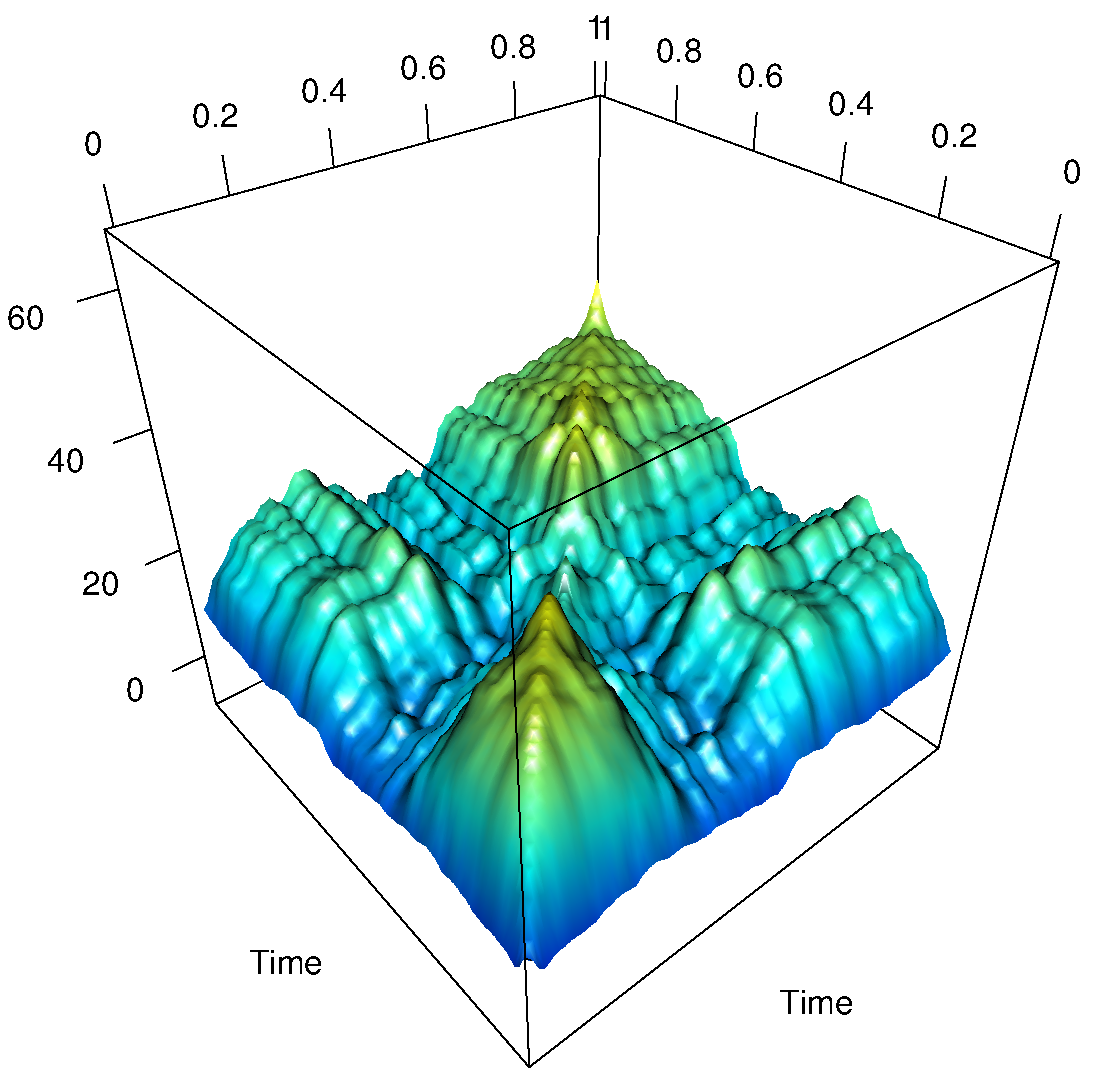}
\includegraphics[height=0.3\textwidth,width=0.3\textwidth]{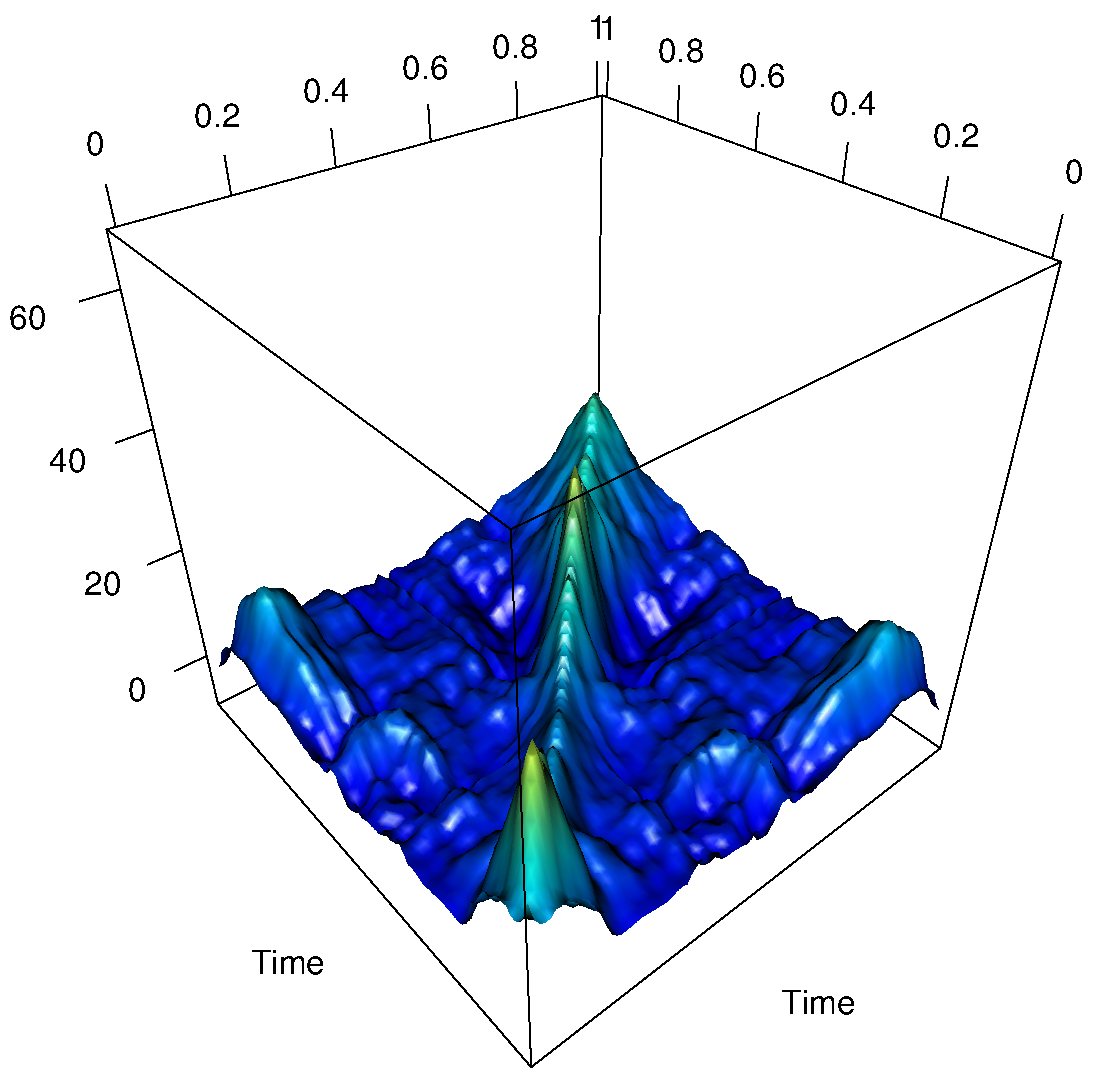}
\includegraphics[height=0.3\textwidth,width=0.3\textwidth]{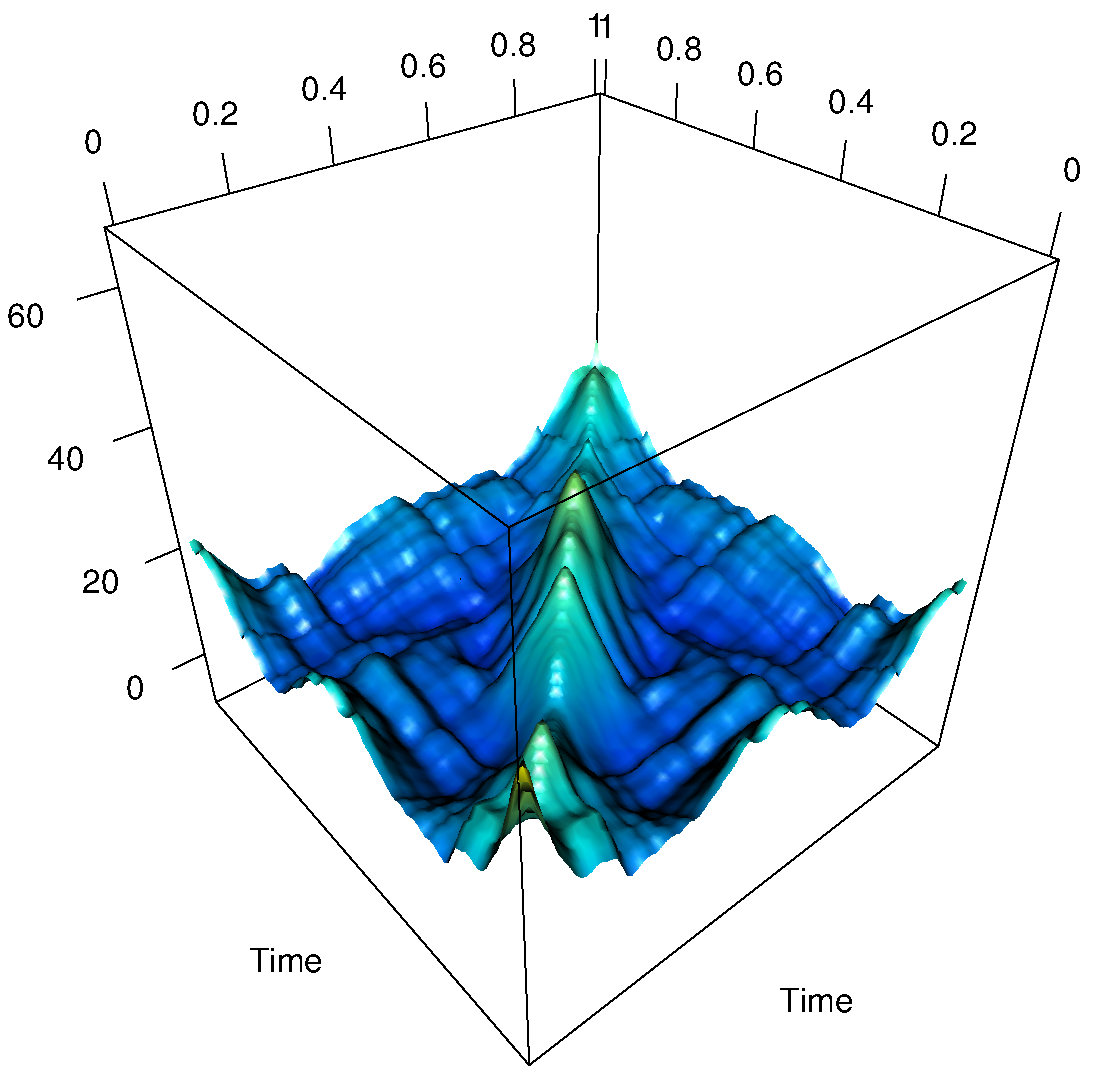}
\caption{Marginal covariance function between times for the five
Romance languages. First row: Italian (left), French (center) and
Portuguese (right). Second row: American Spanish (left) and Iberian
Spanish (right).} \label{fig:lang_time}
\end{figure}

Figures \ref{fig:lang} and \ref{fig:lang_time} show the estimated
marginal covariance functions for the five Romance languages. As can
be seen, the frequency covariance functions present differences
that appear to be language-specific (with peaks and plateau in different positions), while the time covariances
have similar structure, the dependence decreasing when time lag
increases and most of the covariability concentrated close to the diagonal.

\subsection{A permutation test to compare means and covariance operators between groups}
\label{sec:test} We made the assumption above that the covariance
operators are common to all the words within each language, while
the means are different between words. This assumption can be
verified using permutation tests that look at the effect of the
group factor on the parameters of the sound process.

When an estimator for a parameter is available and it is possible to
define a distance $d(.,.)$ between two estimates, a distance-based
permutation test can be set up in the following way. Let
$X_{1l},\ldots,X_{nl}$  be a sample of surfaces from the $l$-th
group under consideration and $K_{l}=K(X_{1l},\ldots,X_{nl})$ be an
estimator for an unknown parameter $\Gamma_l$ of the process which
generates the data belonging to the $l$-th group. In the case of
acoustic phonetic data, this parameter can be for example the mean,
the frequency covariance operator or the time covariance operator.

Permutation tests are non parametric tests which rely on the fact
that, if there is no difference among experimental groups, the group
labelling of the observations (in our case the log-spectrograms) is
completely arbitrary. Therefore, the null hypothesis that the labels
are arbitrary is tested by comparing the test statistic with its
permutation distribution, i.e. the value of the test statistics for
all the possible permutation of labels. In practice, only a subset
of permutations, chosen at random, is used to assess the
distribution. A sufficient condition to apply this permutation
procedure is exchangeability under the null hypothesis. This is
trivially verified in the case of the test for the mean. For the
comparison of covariance operators, this implies the groups having the
same mean. If this not true, we can apply the procedure to the
centred observations $\widetilde{X}_{il}=X_{il}-\overline{X}_{l}$,
$i=1,\ldots,n$, $l=1,\ldots,G$, where $\overline{X}_{l}$ is the
sample mean for the $l$-th group. This guarantees the observations
to be asymptotically exchangeable due to the law of large numbers.

Indeed, if we want to test the null hypothesis that
$\Gamma_1=\Gamma_2=\cdots=\Gamma_G$ against the alternative that the
parameter is different for at least one group,  we can consider as
test statistic
$$
T_0=\frac{1}{G}\sum_{l=1}^G d(K_{l},\overline{K})^2,
$$
where $\overline{K}$ is the sample Fr\'{e}chet mean of
$K_1,\ldots,K_G$, defined as
$$
\overline{K}=\arg \min_{K\in P} \frac{1}{G}\sum_{l=1}^G
d(K_{l},K)^2,
$$
where $P$ is the appropriate functional space to which the parameters
belong. This test statistic measures the variability of the
estimator of the parameters across the different groups. If the
parameter is indeed different for some groups, we expect that the
their estimates from groups $1,\ldots,G$ show greater variability
than those obtained from random permutations of the group labels in
the data set. Thus, large values of $T_0$ are evidence against the
null hypothesis.

Let us take $M$ permutations of the original group
labels and compute the test statistic for the permuted sample
$T_m=\sum_{l=1}^G d(K_l^m,\overline{K}^m)^2$, where $K_l^m$,
$l=1,\ldots,G$ are the estimates of the parameters obtained from the
observations assigned to the group $l$ in the $m$-th permutation and
$\overline{K}^m$ is their sample Fr\'{e}chet mean. The p-value of
the test will therefore be the proportion of permutations for which
the test statistic is greater than in the original data set, i.e.
$p=\frac{\#\{T_m>T_0\}}{M}$.

We apply now this general procedure to the three parameters of
interest in our case, i.e. the mean, the frequency covariance
operator and the time covariance operator, when the groups are the
different words within each language and/or the different language.

Let us start by considering the test to compare the means of the
log-spectrograms across the words (digit) of each language. Here the
natural estimator for the word-wise mean log-spectrogram is the
sample mean, i.e.
$$
K_{l}=\overline{S}_{l}^{L}(\omega,t)=\frac{1}{n_l}\sum_{k=1}^{n_l}
S_{lk}^L(\omega,t)
$$
and the distance can be chosen to be the distance in
$L^2([0,8\ \mathrm{kHz}]\times [0,1])$,
$$
d(\overline{S}_{l}^{L},\overline{S}_{l'}^{L})=\sqrt{\int_0^{8\mathrm{kHz}}\int_0^1
[\overline{S}_{l}^{L}(\omega,t)-\overline{S}_{l'}^{L}(\omega,t)]^2
\mathrm{d} \omega \mathrm{d}t}.
$$
Table \ref{tab:mean} reports the results for the test for the
difference of the means between the digit $l=1,\ldots,10$ for the
five Romance languages. It can be seen that a significant difference
can be found for French and American Spanish and thus we choose
to account for this difference when modelling the sound changes.

We can apply the same procedure to the test for the covariance
operators. First, we need to define a distance between covariance
operators. \citet{Pigoli2014} show that when the covariance operator
is the object of interest for the statistical analysis, a
distance-based approach can be fruitfully used and the choice of the
distance is relevant, different distances catching different
properties of the covariance structure. In particular, they propose a distance based on the geometrical
properties of the space of covariance operators, the
\emph{Procrustes reflection size-and-shape distance}. This distance
uses a map from the space of covariance operators to the space of
Hilbert-Schmidt operators, i.e. compact operator with finite norm
$||L||_{HS}=\mathrm{trace}(L_i^*L_i)$. This being a Hilbert space,
distances between the transformed operators can be easily evaluated.
However, the map is defined up to a unitary operator and a
Procrustes matching is therefore needed to evaluate the distance
between the two equivalence classes. Let $C_1$ and $C_2$ be the
covariance operators we want to compare and $L_1$ and $L_2$ the
Hilbert-Schmidt operators such that $C_i=L_i L_i^*$.
\citet{Pigoli2014} prove that the Procrustes reflection
size-and-shape distance has the explicit analytic expression
$$
d_P(C_1,C_2)^2=||L_1||_{HS}^2+||L_2||_{HS}^2-2\sum_{k=1}^{\infty}
\sigma_k,
$$
where $\sigma_k$ are the the singular values of the compact operator
$L_2^*L_1$. A possible map is the square root $L_i=(C_i)^{1/2}$ (although the distance itself is invariant to the choice of map) and
we use this choice in the following analysis, where we analyze the
five selected Romance languages looking at the Procrustes distance
between their frequency covariance operators.

For a given choice of the distance, a sample Fr\'{e}chet mean and
variance of a set of covariance operators $C^1,\ldots,C^L$ can be
defined as
$$
\overline{C}=\arg \inf_C \sum_{L=1}^G d(C^L,C)^2, \hspace{10mm}
\widehat{\sigma}^2= \frac{1}{G} \sum_{L=1}^G d(C^L,\overline{C})^2.
$$
These provide estimates for the centre point and the variability of
the distribution with respect to the distance $d(.,.)$, which are
needed for the test statistic in the permutation test.

\begin{table}
\caption{P-values of the permutation tests for H$0$:
$\mu_1^L=\mu_{2}^L=\cdots=\mu_{10}^L$ \emph{vs} H$1$: at least one
is different, where $\mu_{i}^L$ is the mean log-spectrogram for the
language $L$ and word $i$, for the five Romance
languages.\label{tab:mean}}\vspace{0.2cm}
\begin{tabular}{|r|c|c|c|c|c|}
\hline
Language& French   &  Italian &  Portuguese & American Spanish & Iberian Spanish\\
    \hline
$p-value$ &  $<$0.001& 0.02& 0.96& $<$0.001& 0.205\\
\hline
\end{tabular}
\end{table}

\begin{table}
\caption{P-values of the permutation tests for H$0$:
$C_{\omega,1}^L=C_{\omega,2}^L=\cdots=C_{\omega,10}^L$ \emph{vs}
H$1$: at least one is different, where $C_{\omega,i}^L$ is the
marginal frequency covariance operator for the language $L$ and word
$i$, for the five Romance languages. The Procrustes distance is used
for the test statistic.\label{tab:test}}\vspace{0.2cm}
\begin{tabular}{|r|c|c|c|c|c|}
\hline
Language& French   &  Italian &  Portuguese & American Spanish & Iberian Spanish\\
    \hline
$p-value$&  0.113& 0.991& 0.968& 0.815& 0.985\\
\hline
\end{tabular}
\end{table}

\begin{table}
\caption{P-values of the permutation tests for H$0$:
$C^L_{t,1}=C^L_{t,2}=\cdots=C^L_{t,10}$ \emph{vs} H$1$: at least one
is different, where $C_{t,i}^L$ is the marginal time covariance
operator for the language $L$ and word $i$, for the five Romance
languages. The Procrustes distance is used for the test
statistic.\label{tab:testt}}\vspace{0.2cm}
\begin{tabular}{|r|c|c|c|c|c|}
\hline
Language& French   &  Italian &  Portuguese & American Spanish & Iberian Spanish\\
    \hline
$p-value$&  0.02 & 0.422&  0.834& 0.683& 0.17\\
\hline
\end{tabular}
\end{table}

Using this procedure, we can verify whether the assumption that the
covariance operators are the same across the words is disproved by
the data. Table \ref{tab:test} shows the p-values of the permutation
tests for the equality of the marginal frequency covariance operator
across the different words for the five Romance languages described
in Section \ref{sec:data}, obtained with the Procrustes distance between
sample covariance operators and $M=1000$ permutations on the
residual log-spectrograms. In the interpretation of these p-values, we need to account for the multiple tests that have been carried out. By applying a Bonferroni correction to the p-values in Table  \ref{tab:test},  it can be seen that there is no evidence
against the hypothesis that the covariance operator is the same for
all words for all the considered languages. The same is true for the time covariance operator, as
can be seen in Table \ref{tab:testt}, which reports the unadjusted p-values of this second test.

A possible concern is that the dimension of the data set becomes
relatively small when it is split between the different words and
language and therefore these testing procedure will have little
power. On the other hand, this reasoning encourages us to simplify the
model (assuming covariance operators constant across words) so that
enough observations are available to estimate the parameters
accurately. With a larger data set that allows us to highlight
differences between word-wise covariance operators, we would have more information to estimate these operators accurately.

\section{Exploring phonetic differences}
\label{sec:path}

We now have the tools to explore the phonetic differences between the
languages in the Oxford Romance languages data set. This can be done at different levels. A possible way to go would be to
pair two speakers belonging to two different languages and look at
their difference. However, this neglects the variability of the
speech within the language and it would not be clear which aspects of
the phonetic differences are to be credited to the difference between
languages and which to the difference between the two individual
speakers, unless we had available recordings from bilingual
subjects. In this section we present a possible approach to the
modelling of phonetic changes that takes into account the features of
the speaker's population.

\subsection{Modelling changes in the parameters of the phonetic process}
We can start by looking at the path that links the mean of the
log-spectrograms between two words of different languages. These
should be two words known to be related in the languages'
historical development. This is the case for example for the same digit in any two
different Romance languages.

Considered as functional objects, the log-spectrograms means are
unconstrained and integrable surfaces, thus interpolation and
extrapolation can be simply obtained with a linear combination,
where the weights are determined from the distance of the language
we want to predict from the known languages. For example, if we want
to reconstruct the path of the mean for the digit $i$ from the
language $L_1$ to the language $L_2$, we have
$$
\overline{S}_i(x)=\overline{S}_i^{L_1}+x(\overline{S}_i^{L_2}-\overline{S}_i^{L_1}),
$$
where $x\in[0,1]$ provides a linear interpolation from language
$L_1$ to language $L_2$, while $x<0$ or $x>1$ provides an
extrapolation in the direction of the difference between the two
languages, with $\overline{S}_i^{L}$ being the mean of the
log-spectrograms from speakers of the language $L$ pronouncing the
$i$-th digit. For example,  Figure \ref{fig:sound_mean} shows six steps along a
reconstructed path for the mean log-spectrogram of ``one'', from French [\~\oe] to
Portuguese [\~ u]. Indeed, this path has historical significance, as the sound change from Latin ``unus'' to French ``un'' likely went via the sound [\~ u], which is still maintained in modern Portuguese (it should be noted that we are, of course, not implying that modern French is derived from Portuguese, but merely that a historical sound of modern French is maintained in Portuguese).

A natural question is whether this can replicated for the covariance
structure, in order to interpolate and extrapolate a more general
description of the sound generation process. However, the case of
the covariance structure is more complex. Experience with low
dimensional covariance matrices \citep[see][]{Dryden2009} and the
case of the frequency covariance operators illustrated in
\citet{Pigoli2014} show that a linear interpolation is not a good
choice for objects belonging to a non-Euclidean space. We want
therefore to use a geodesic interpolation based on an appropriate
metric for the covariance operator. Moreover, since we model the
covariance structure as separable, we also want  the predicted
covariance structure to preserve this property. It is not possible
to do this with geodesic paths in the general space of
four-dimensional covariance structures and thus we define the new
covariance structure as the tensor product of the geodesic
interpolations (or extrapolations) in the space of time and
frequency covariance operators,
$$
C^x=C_{\omega}^x \otimes C_{t}^x,
$$
where the geodesic interpolations (or extrapolations)
$C_{\omega}^x$, $C_{t}^x$ depend on the chosen metric. In the case
of the Procrustes reflection size and shape distance, the geodesic
has the form
$$
C_r^x=[(C_{r}^{L_1})^{1/2}+x((C_{r}^{L_2})^{1/2}\widetilde{R}-(C_{r}^{L_1})^{1/2})][(C_{r}^{L_1})^{1/2}+x((C_{r}^{L_2})^{1/2}\widetilde{R}-(C_{r}^{L_1})^{1/2})]^*
$$
where $r=\omega, t$ and $\widetilde{R}$ is the unitary operator
that minimizes
$||(C_{r}^{L_1})^{1/2}-(C_{r}^{L_2})^{1/2}R||^2_{HS}$
\citep[see][]{Pigoli2014}. Other choices of the metric are of course
possible, as long as they provide a valid geodesic for the
covariance operator. However, some preliminary experiments reported
in \citet{Pigoli2014} suggest that the Procrustes reflection size
and shape geodesic performs better in the extrapolation of frequency
covariance operators than existing alternatives.

\subsection{What would someone sound like speaking in a different language?}

The framework we have set up allows us also to observe how the sound
produced by a speaker would be modified as we move to a different
language. As mentioned in the introduction, we aim to map the
sound produced by this speaker to that of a hypothetical speaker
with the same position in the space of possible speakers in a different language, with respect to the
language variability structure. To do this, we need some additional
specification of the statistical model which generates the
log-spectrograms. For example, if we assume that the
log-spectrograms of a spoken word are generated from a Gaussian
process, its distribution is fully determined by the mean
log-spectrogram (which is expected to be word-dependent) and the
covariance structure. More generally, we identify the population of possible
pronunciations of a specific word of a language through its mean
log-spectrogram, which is word-specific, and its time and frequency
covariance functions, which are properties of the whole language.
Thus, we identify as a speaker-specific residual what is left in
the phonetic data once means and covariance information have been
removed. Let us denote with $\mathfrak{F}_i^L$ this operation for
the word $i$ of the language $L$. Then, we can obtain a
representation of the log-spectrogram for a speaker from a language
$L_1$ in the language $L_2$ as
\begin{equation}
\label{eq:proj} S_{ik}^{L_1\rightarrow
L_2}=[\mathfrak{F}_i^{L_2}]^{-1}\circ \mathfrak{F}_i^{L_1}
(S_{ik}^{L_1}). \end{equation}
We choose to use the same word
for both languages because in our data set words can be
paired in a sensible way (the various pronunciations of the same digit in two Romance languages
sharing a common historical origin).

\begin{figure}[h!]
\includegraphics[scale=0.28]{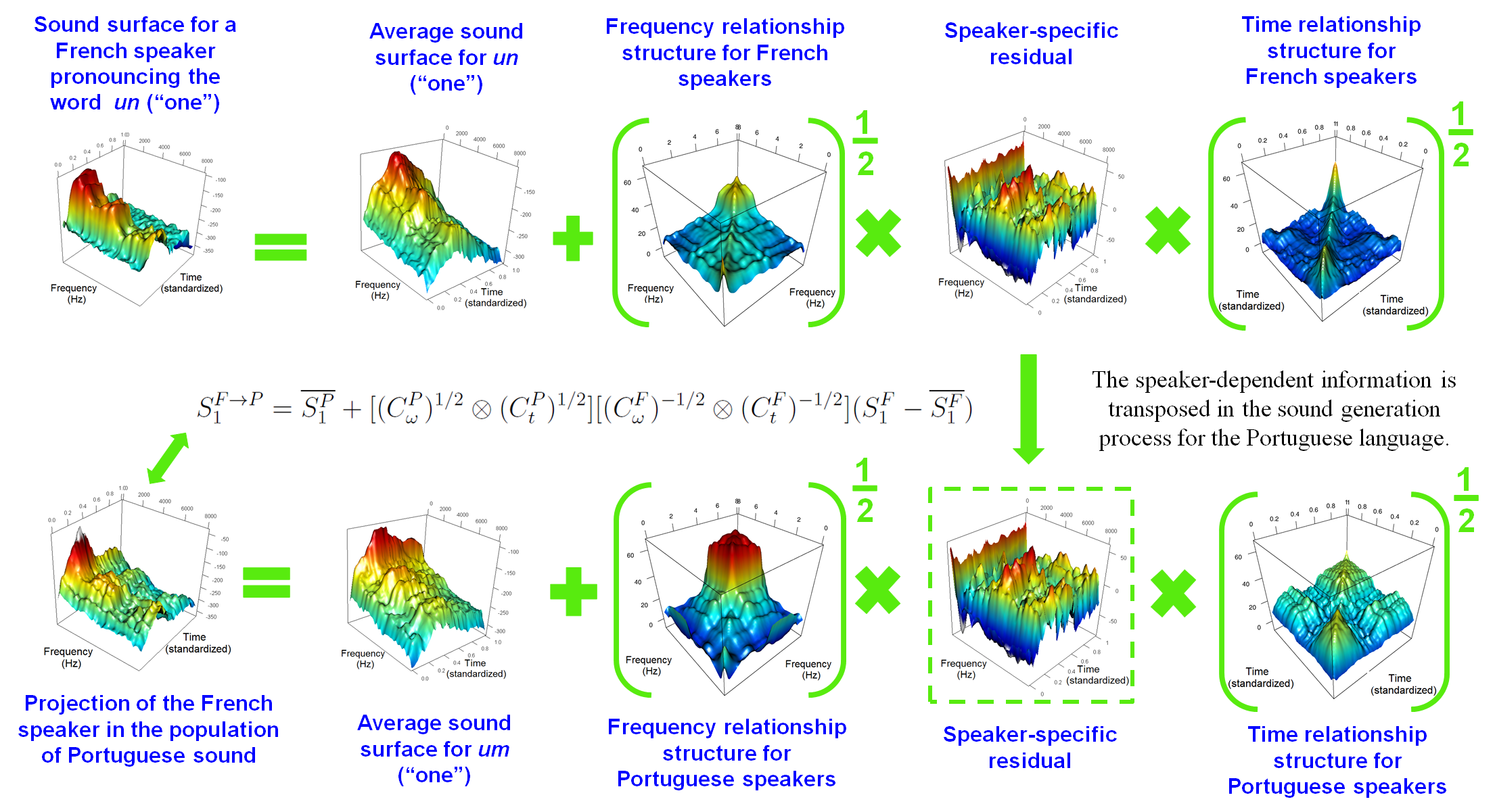}
\caption{Example of the mapping of a French speaker's
log-spectrogram to the same position in the space of Portuguese
pronunciations fo the corresponding word.}\label{fig:example}
\end{figure}

The challenge now is how to define the transformation
$\mathfrak{F}_i^L$.  This is obtained considering both the
characteristics of the sound populations in the two languages and
the relative ``position'' of the speaker in their language vis-a-vis all the other speakers. A
graphical representation of this idea for the case of a French
speaker mapped to the Portuguese language can be seen in Fig.
\ref{fig:example}. To define this transformation, we start from a
speaker $k$ from the language $L_1$ and we consider the residual
log-spectrogram $R_{ik}^{L_1}=S_{ik}^{L_1}-
\overline{S}_{i.}^{L_1}$. We would like to apply now a
transformation that makes this residual uncorrelated, as generated
by a white noise process. Let us consider the transformation from a
finite dimensional white noise
$$
Z=\sum_{i,j}^p z_{ij} v_i^{\omega}\otimes v^t_j, \hspace{5mm}
z_{ij}\sim N(0,1)
$$
to a random surface with the same mean and covariance structure
$(C^{L_1}_{\omega})^{1/2}_1\otimes (C^{L_1}_t)^{1/2} Z$ of the sound
distribution. We use here the notation for the application of a
tensorised operator where
$$
L_1\otimes L_2  Z (\omega,t)=\int\int
l_1(\omega,y)z(x,y)l_2(x,t)\mathrm{d}x\mathrm{d}y.
$$
To obtain $\mathfrak{F}_i^L$, we would need to invert the
transformation from $Z$ to the sound process. This is not possible
in general (due to the unbounded nature of inverse covariance operators), but we can restrict the inverse to work on the subspaces
spanned by our data, thus defining $(C^L_l)^{-1/2}=\sum_{j=1}^{N}
(\lambda_j)^{-1/2}\phi_j \otimes \phi_j$, $\phi_j,j=1,\ldots,N$,
$\{\lambda_j, \phi_j\}$ being eigenvalues and eigenfunctions for
$C^L_l$. We then obtain
$$
\mathfrak{F}_i^L (S_{ik}^{L})=(C^{L}_{\omega})^{-1/2}\otimes
(C^{L}_t)^{-1/2} (S_{ik}^{L}- \overline{S}_{i.}^{L})
$$
and
$$
[\mathfrak{F}_i^L]^{-1} (Z)=(C^{L}_{\omega})^{1/2}\otimes
(C^{L}_t)^{1/2} Z+ \overline{S}_{i.}^{L}.
$$

\begin{figure}[h!]
\includegraphics[height=0.35\textwidth,width=0.32\textwidth]{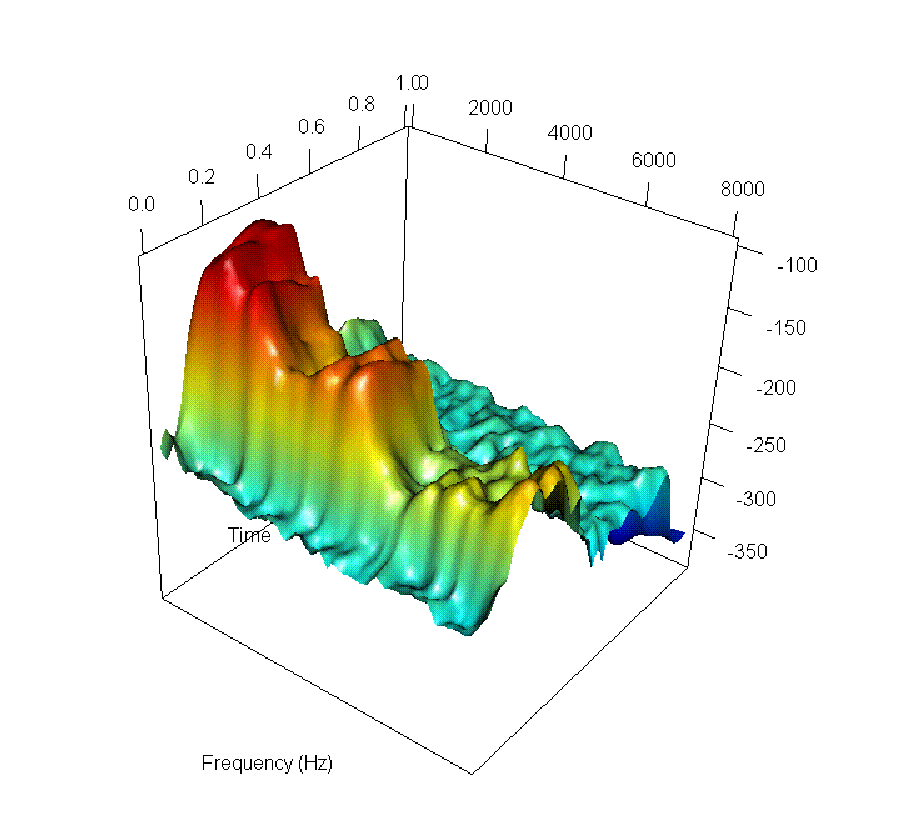}
\includegraphics[height=0.35\textwidth,width=0.32\textwidth]{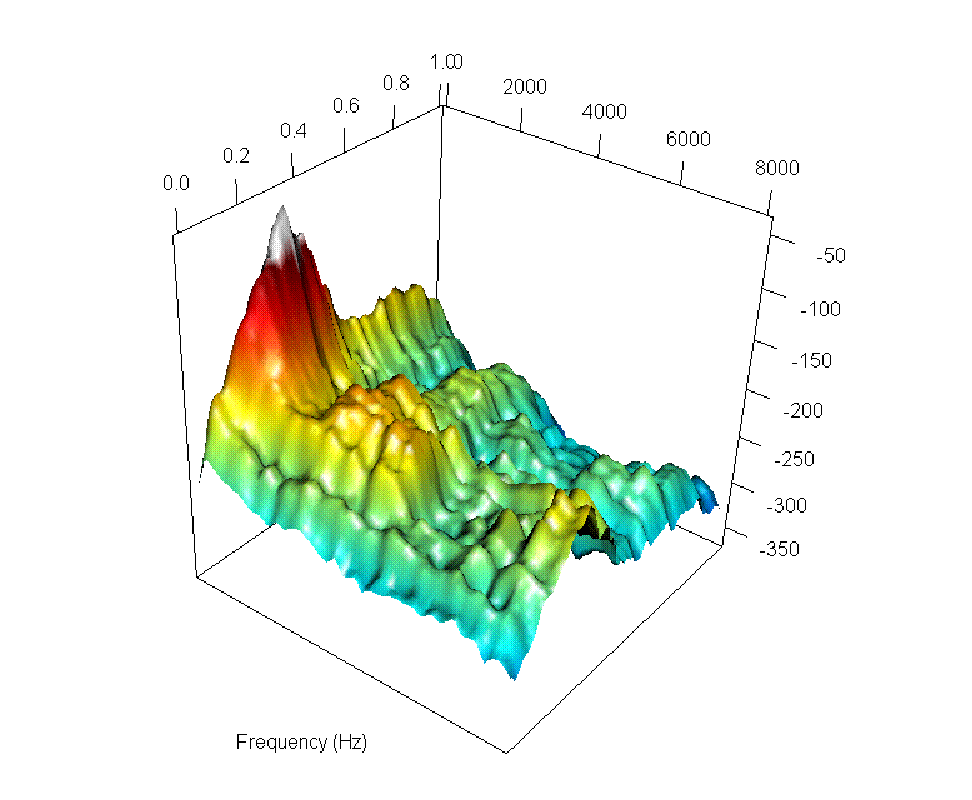}
\includegraphics[height=0.35\textwidth,width=0.32\textwidth]{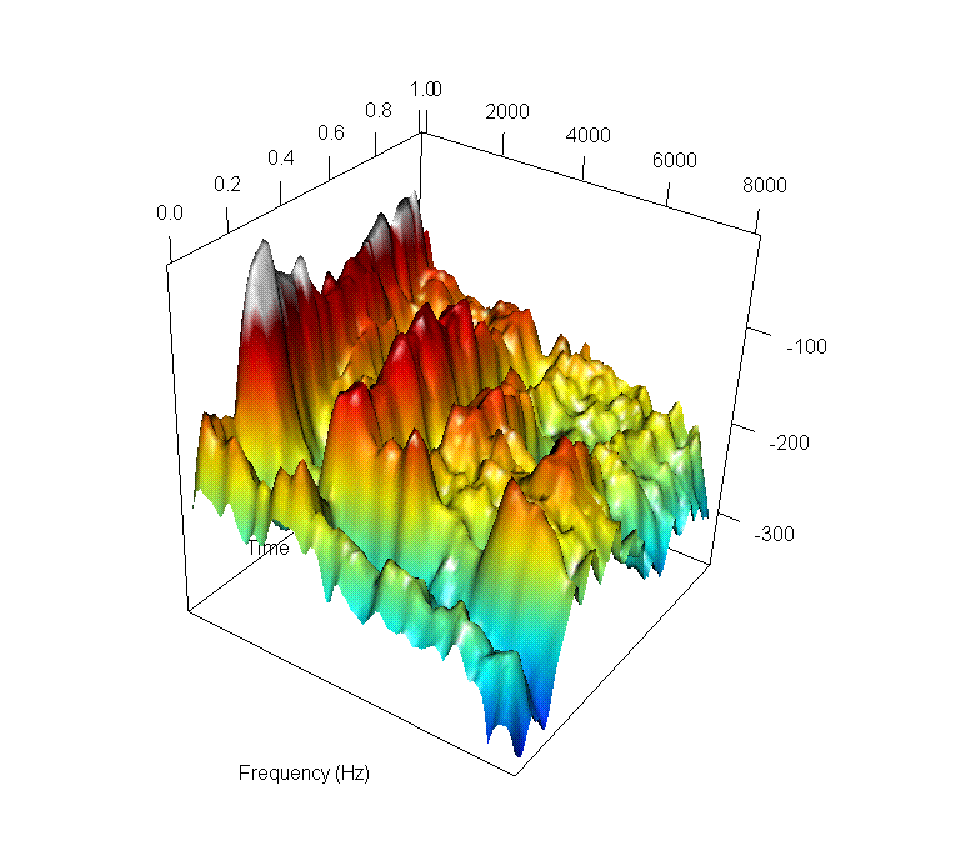}
\caption{Log-spectrograms for the word \textit{un} (``one'') as spoken by a French speaker
(left), its representation as word  \textit{um} (``one'') in Portuguese using
equation (\ref{eq:proj}) (center) and the closest observed word
\textit{um} (``one'') spoken by a Portuguese speaker. }\label{fig:proj}
\end{figure}

\begin{figure}[h!]
\includegraphics[height=0.35\textwidth,width=0.32\textwidth]{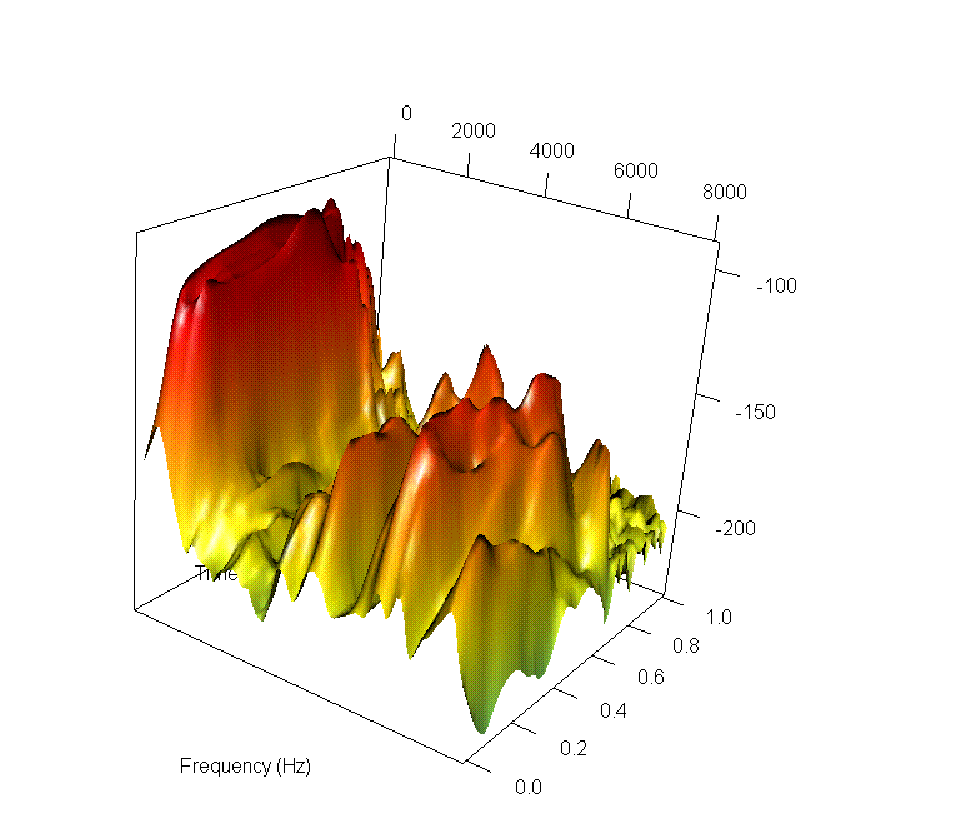}
\includegraphics[height=0.35\textwidth,width=0.32\textwidth]{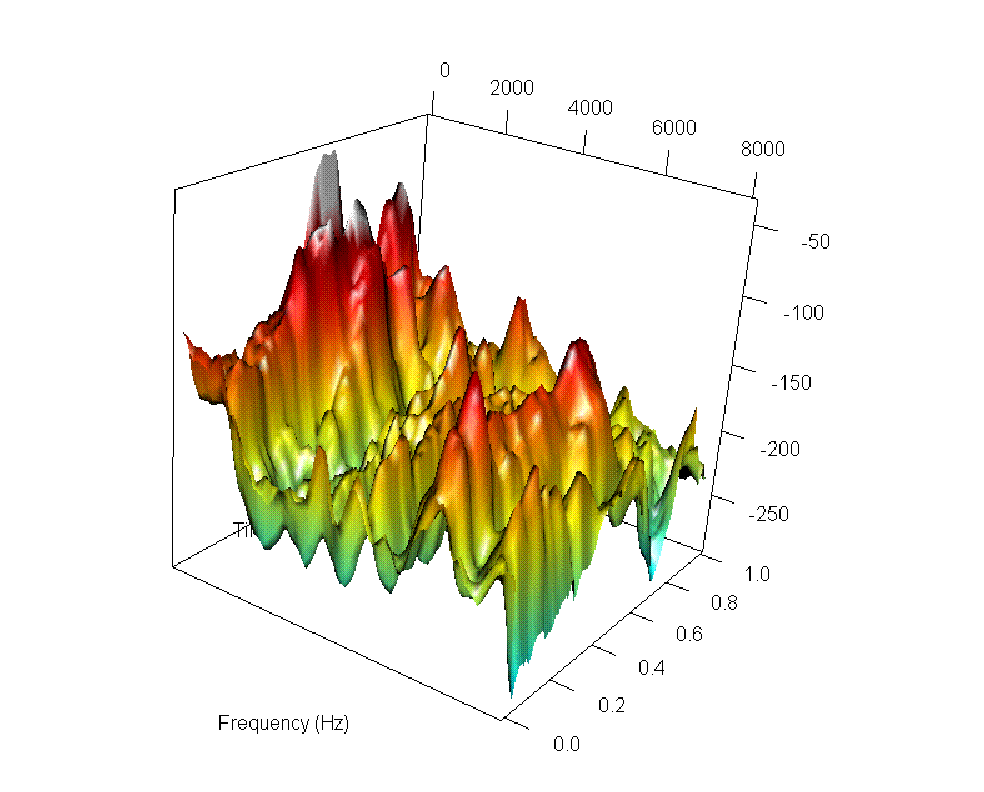}
\includegraphics[height=0.35\textwidth,width=0.32\textwidth]{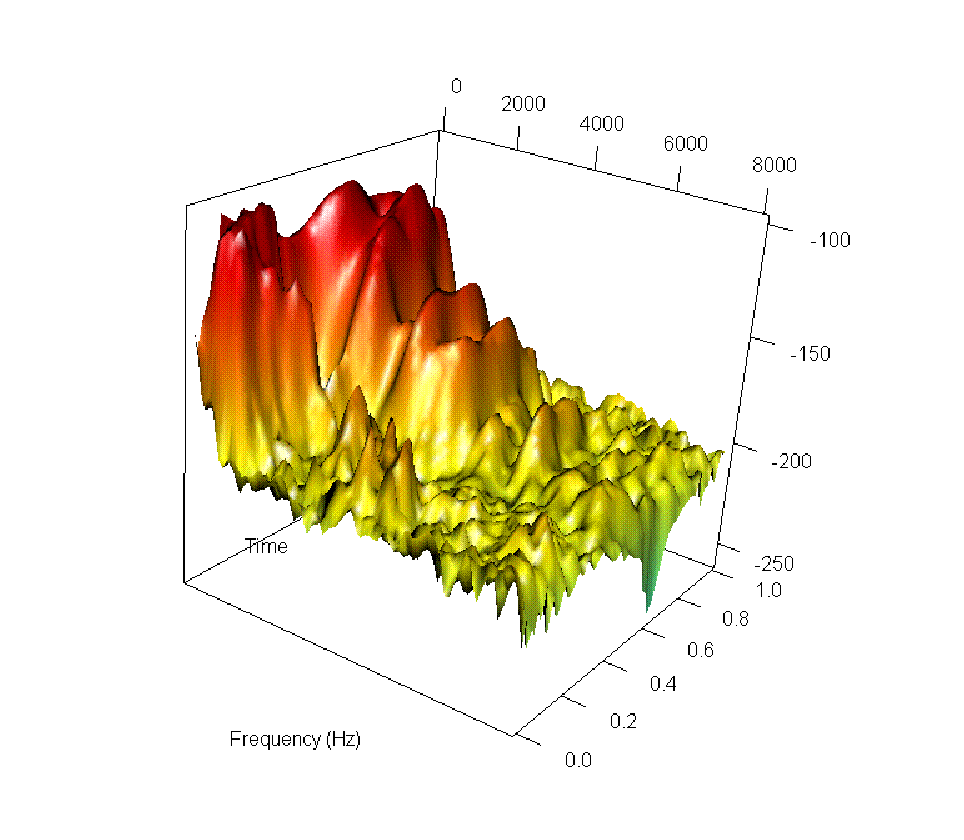}
\caption{Log-spectrograms for the word \textit{uno} (``one'') as spoken by an Italian
speaker (left), its representation as word  \textit{uno} (``one'') in Spanish using
equation (\ref{eq:proj}) (center) and the closest observed word
\textit{uno} (``one'') spoken by a Spanish speaker. }\label{fig:proj2}
\end{figure}

Figure \ref{fig:proj} shows the log-spectrograms for the word \textit{un} (``one'')
of the first French speaker $S_{11}^{Fr}$, its representation when mapped to
Portuguese \textit{um} (``one'') $S_{11}^{Fr\rightarrow P}$ and the closest
observed instance of Portuguese \textit{um} as spoken by a Portuguese speaker, while Fig.\ \ref{fig:proj2} reports the
result of the same operation applied to an Italian speaker, transforming Italian \textit{uno} (``one'') into
Iberian Spanish \textit{uno} (``one''). Though the spelling is the same in this case, the pronunciation of the word in the two languages is not identical, albeit similar.

\subsection{Interpolation and extrapolation of spoken phonemes}
\label{sec:extra} 
\begin{figure}[h!]
\includegraphics[height=0.3\textwidth,width=0.3\textwidth]{F.png}
\includegraphics[height=0.3\textwidth,width=0.3\textwidth]{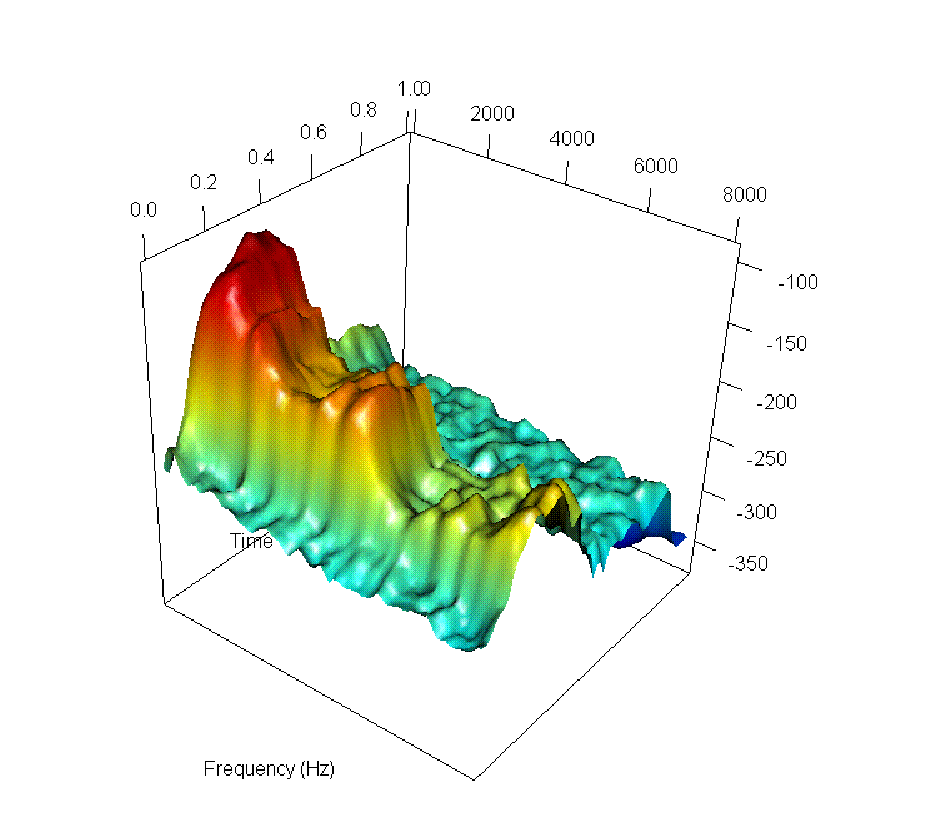}
\includegraphics[height=0.3\textwidth,width=0.3\textwidth]{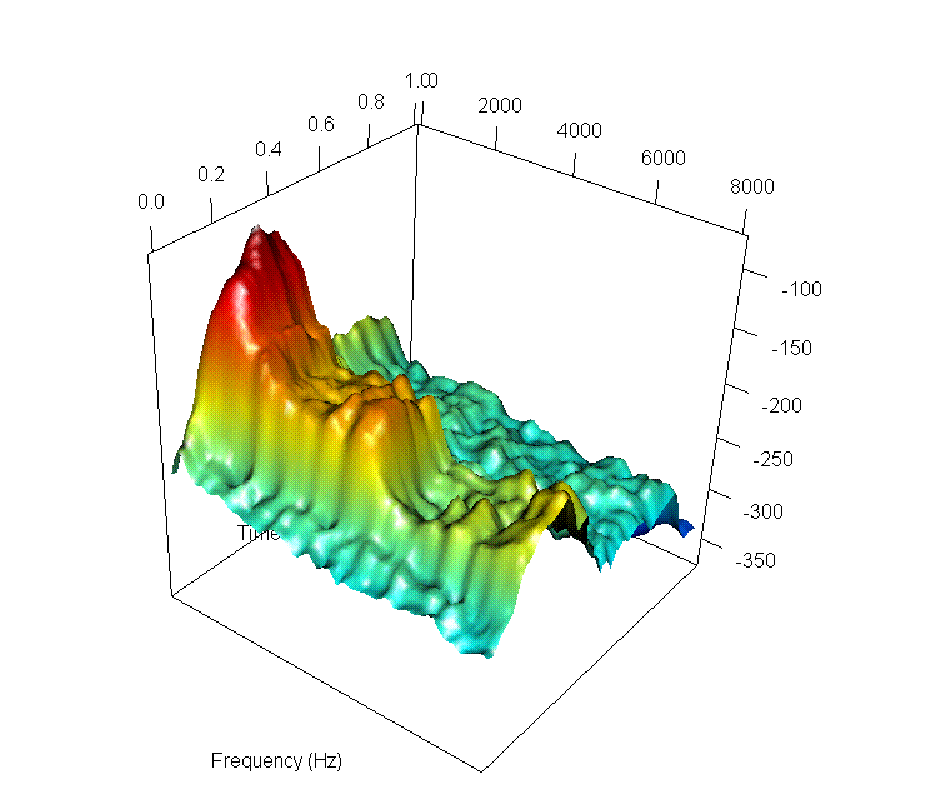}\\
\includegraphics[height=0.3\textwidth,width=0.3\textwidth]{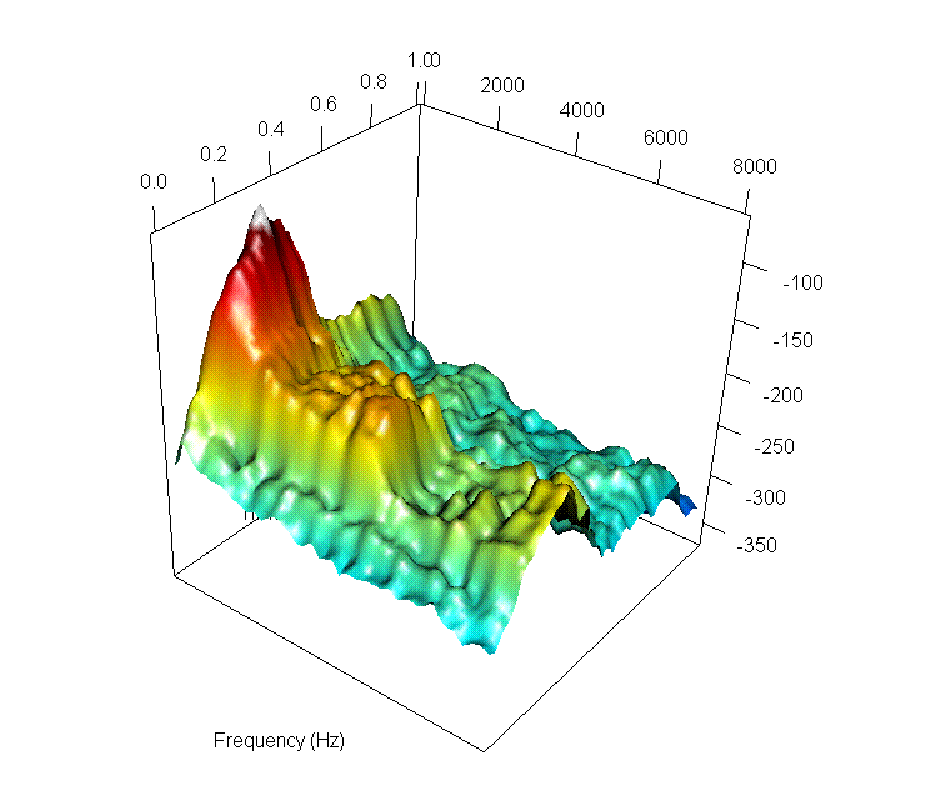}
\includegraphics[height=0.3\textwidth,width=0.3\textwidth]{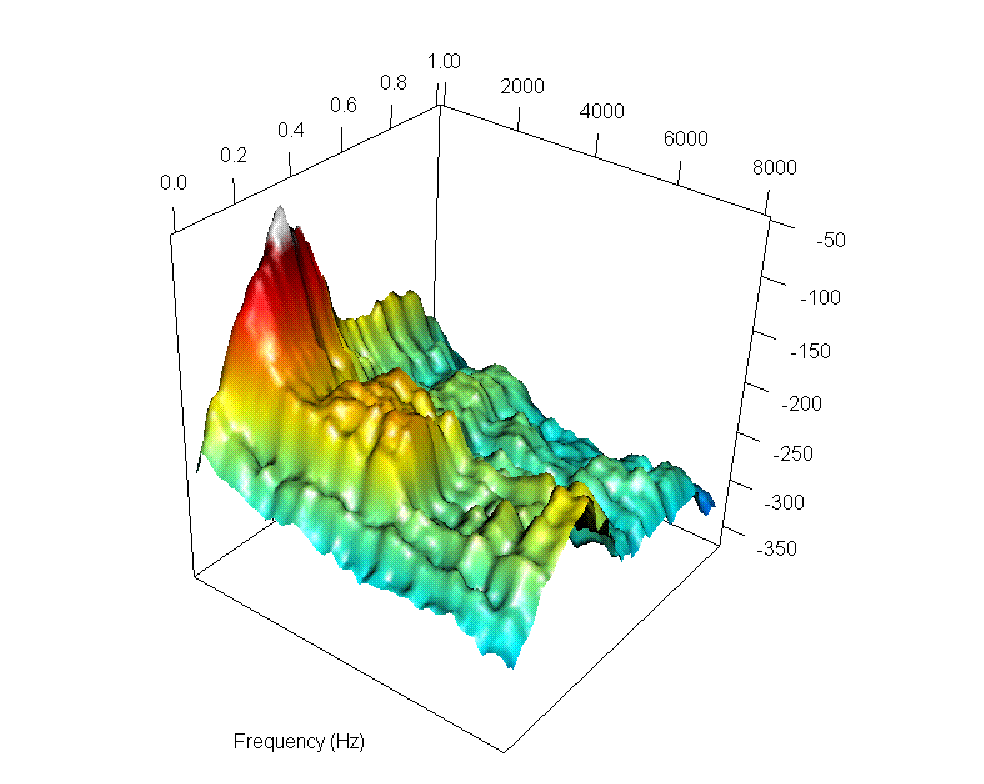}
\includegraphics[height=0.3\textwidth,width=0.3\textwidth]{Fr_P.png}
\caption{Six steps along the smooth path between the log-spectrogram for the word  \textit{un} (``one'')
as spoken by a French speaker (top left) and its representation in Portuguese
(bottom right).}\label{fig:sound_morph}
\end{figure}

The representation of a speaker as they would sound when speaking another language is interesting but is not enough for scholars to
explore the historical sequence of changes that occurred between two languages: a smooth
estimate of the path of change is needed. This is also the case if it is desired to
extrapolate the sound transformation process beyond the path connecting the two
languages, which we recall is a main goal of the Ancient
Sounds project. Luckily, we can use the interpolated means and
covariance operators described above to characterize the unobserved possible
languages that are the intermediate steps in the phonetic
path between two given languages. We thus obtain a smooth path between $S_{ik}^{L_1}$ and its
representation in the language $L_2$ as
\begin{equation}
\label{eq:interp} S_{ik}^{L_1\rightarrow L_2}(x)=[\mathfrak{F}_i^{x}]^{-1}\circ
\mathfrak{F}_i^{L_1} (S_{ik}^{L_1}),
\end{equation}
$[\mathfrak{F}_i^{x}]^{-1}=(C_{\omega}(x))^{1/2}\otimes
(C_t(x))^{1/2} Z+ M(x)$, where $C_{\omega}(x)$ is the interpolated
(or extrapolated) frequency covariance operator, $C_t(x)$ the
correspondent time covariance operator and $M(x)$ the word-dependent
mean. An example of a smooth path between the log-spectrogram for
the word \textit{un} as spoken by the same French speaker considered in the
previous section and its corresponding acoustic representation in Portuguese can be seen in
Fig.\ \ref{fig:sound_morph}.

\begin{figure}[h!]
\includegraphics[height=0.3\textwidth,width=0.3\textwidth]{F.png}
\includegraphics[height=0.3\textwidth,width=0.3\textwidth]{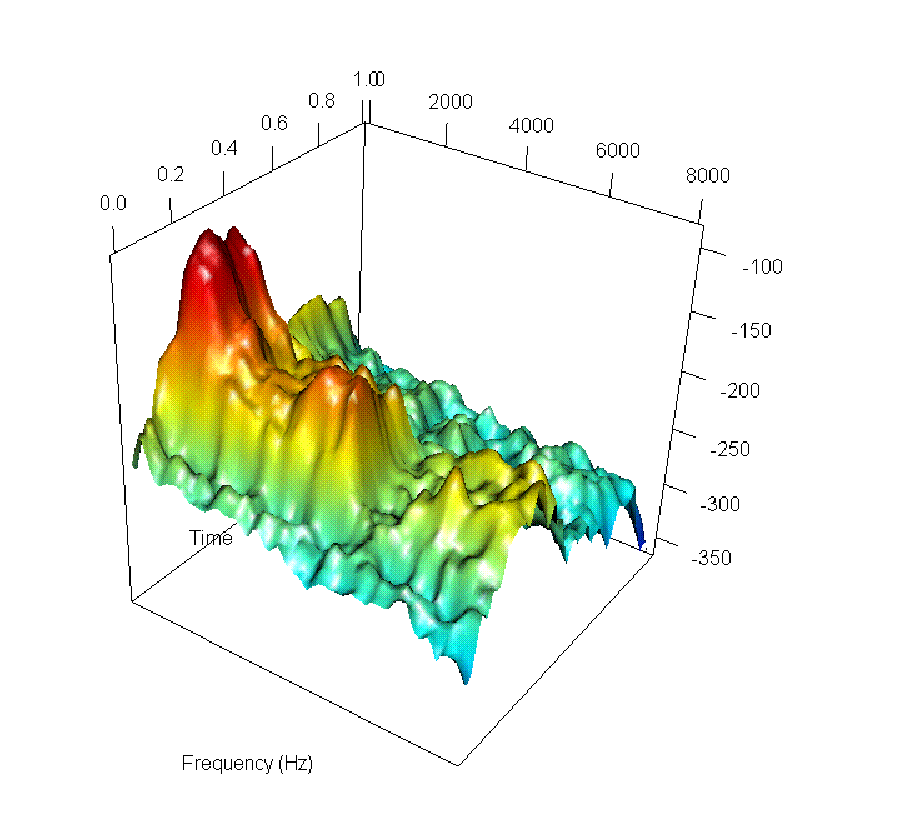}
\includegraphics[height=0.3\textwidth,width=0.3\textwidth]{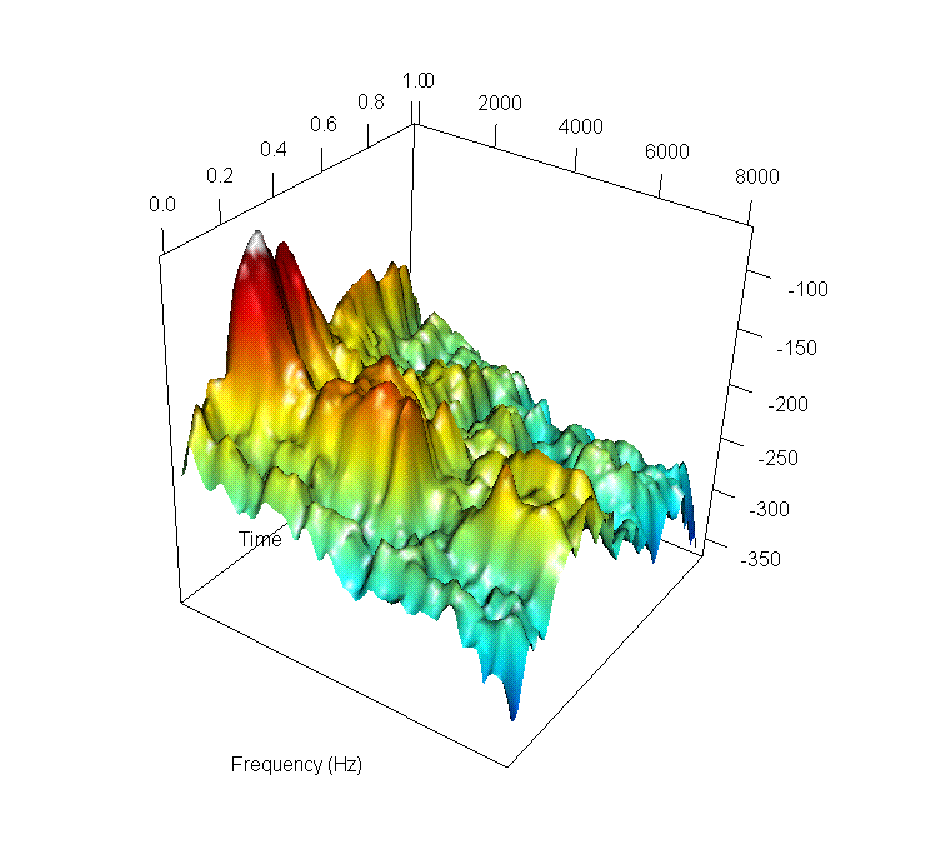}\\
\includegraphics[height=0.3\textwidth,width=0.3\textwidth]{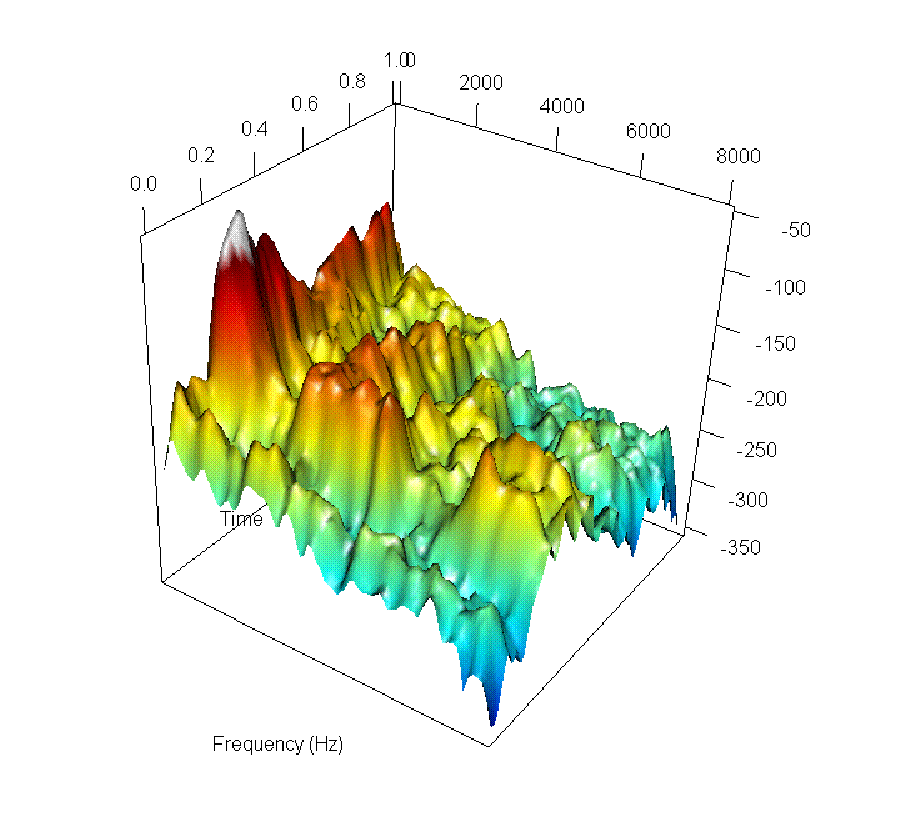}
\includegraphics[height=0.3\textwidth,width=0.3\textwidth]{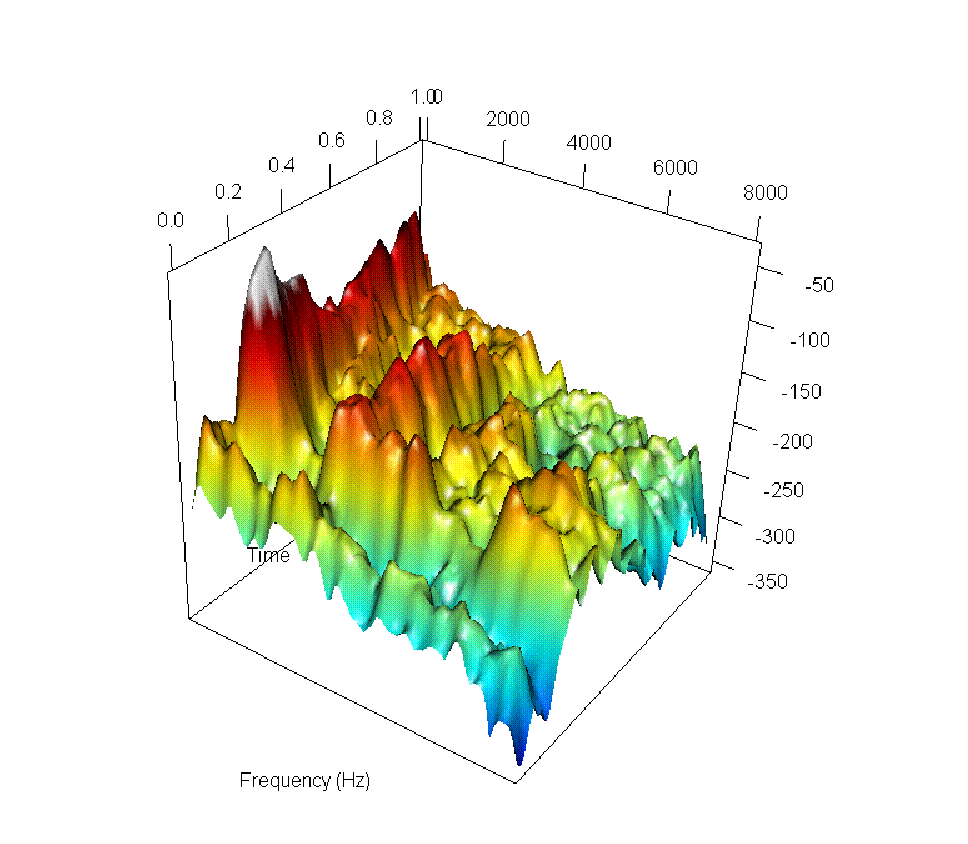}
\includegraphics[height=0.3\textwidth,width=0.3\textwidth]{P.png}
\caption{Six steps along the smooth path between the log-spectrograms for the word
 \textit{un} (``one'') as spoken by a French speaker (top left) and the one for the word
 \textit{um} (``one'') closest to its  transformed representation in Portuguese (bottom
right).}\label{fig:sound_res}
\end{figure}

\begin{figure}[h!]
\includegraphics[height=0.3\textwidth,width=0.3\textwidth]{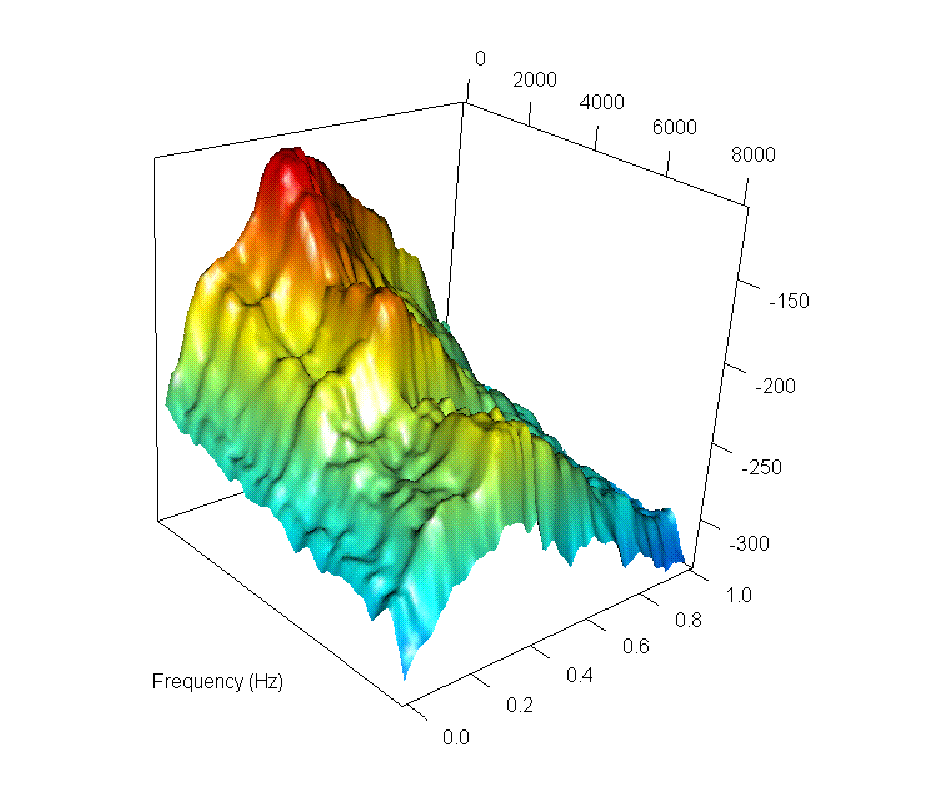}
\includegraphics[height=0.3\textwidth,width=0.3\textwidth]{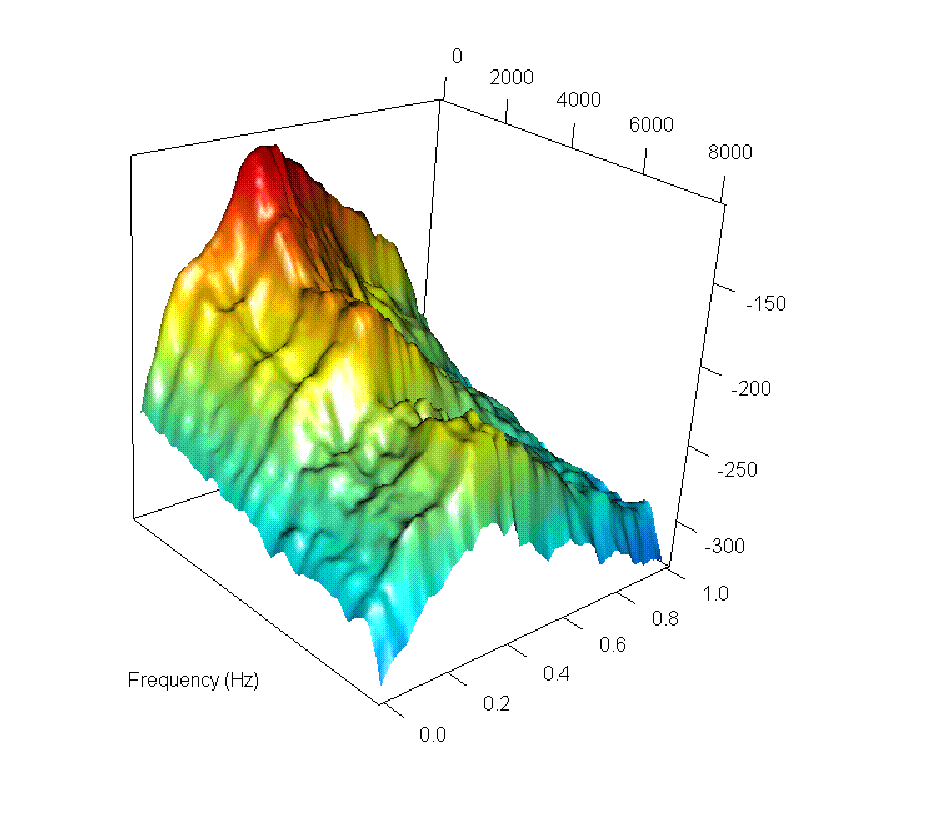}
\includegraphics[height=0.3\textwidth,width=0.3\textwidth]{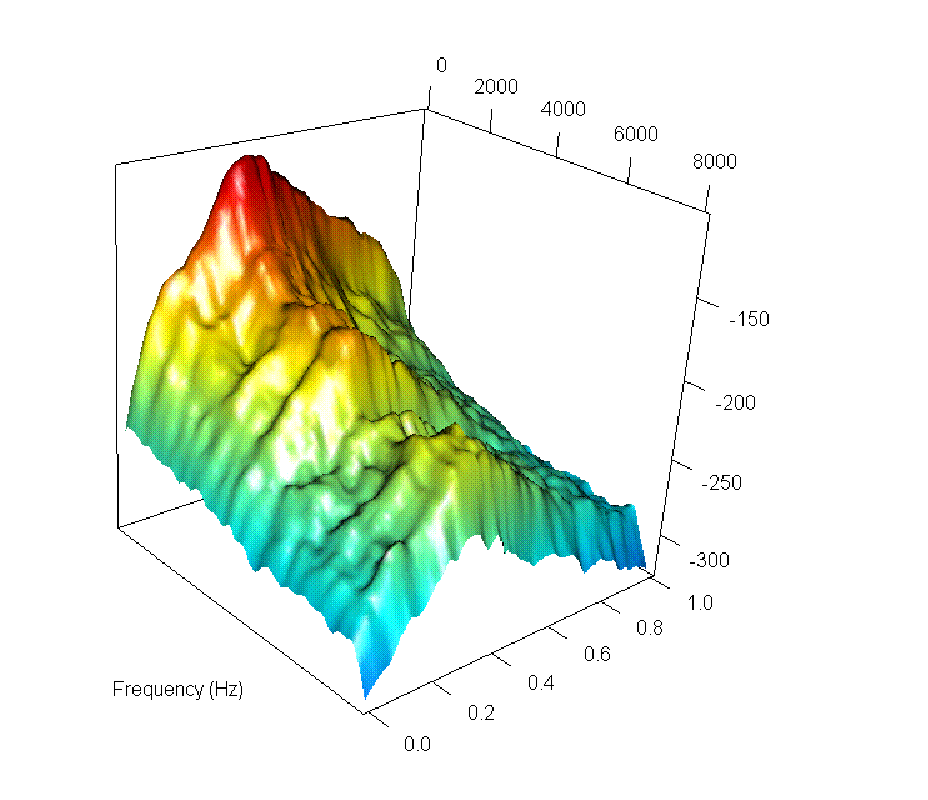}\\
\includegraphics[height=0.3\textwidth,width=0.3\textwidth]{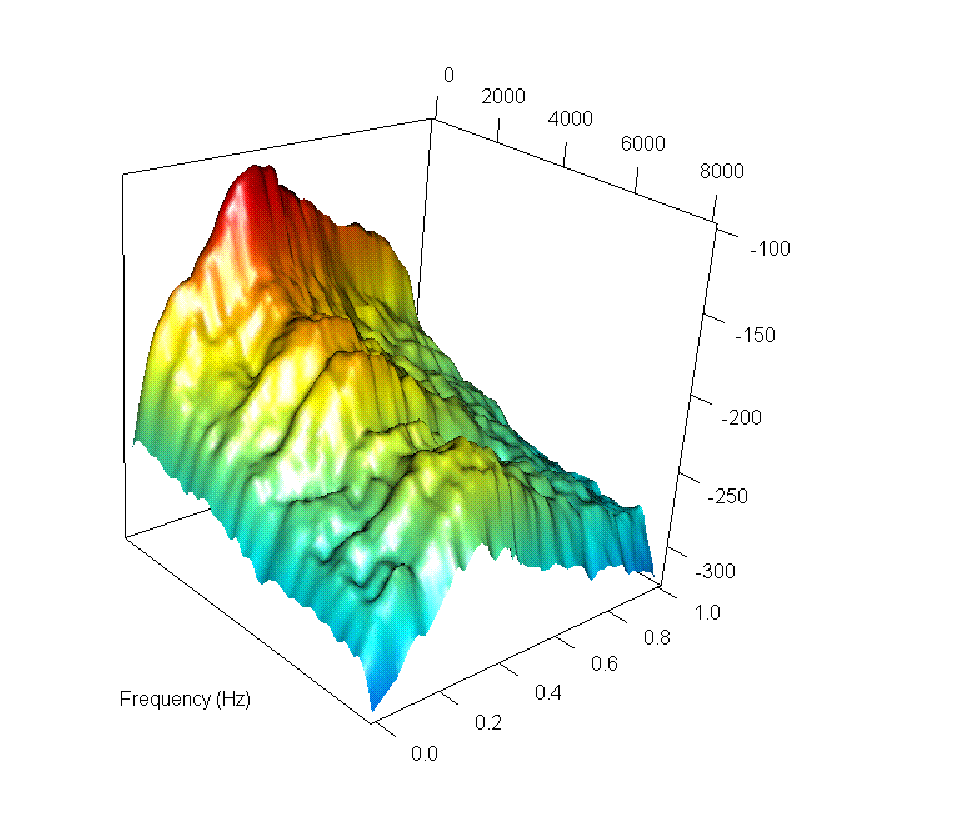}
\includegraphics[height=0.3\textwidth,width=0.3\textwidth]{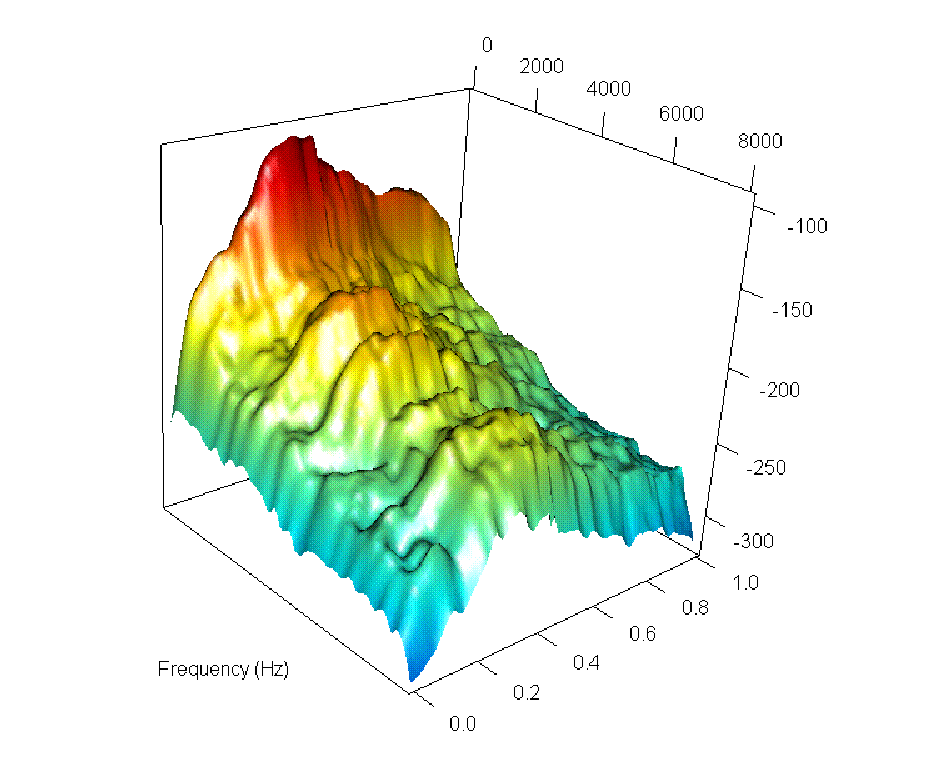}
\includegraphics[height=0.3\textwidth,width=0.3\textwidth]{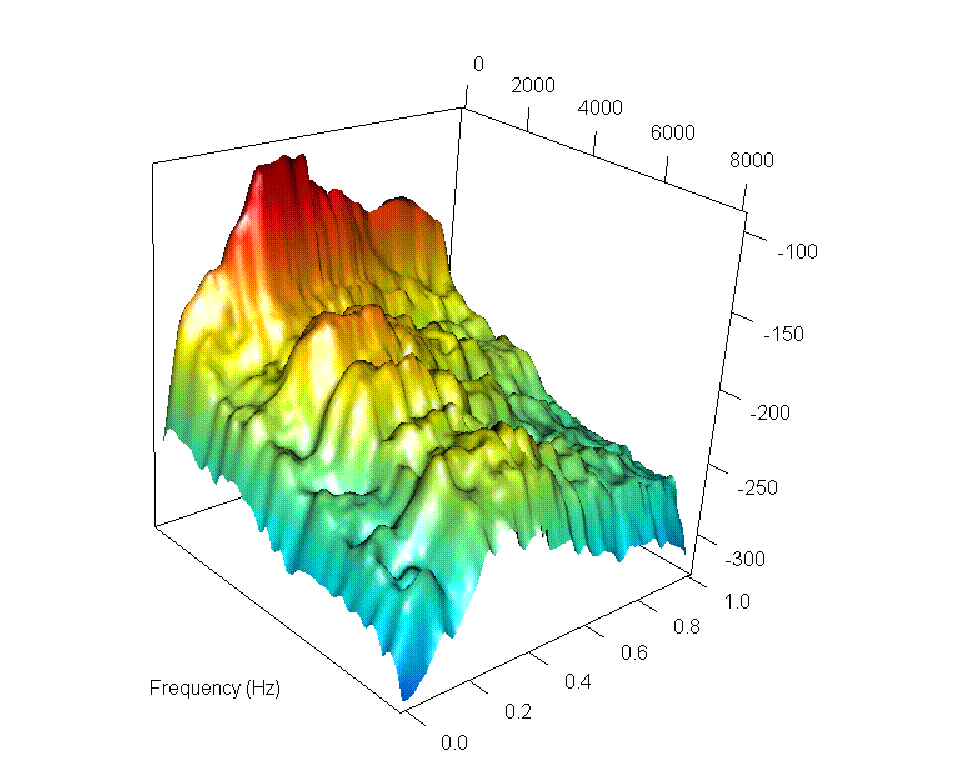}
\caption{Six steps along the smooth path $M(x)$ between the mean log-spectrogram for the
word  \textit{un} (``one'') in French  (top left) and the one for the word  \textit{um} (``one'') in
Portuguese (bottom right).}\label{fig:sound_mean}
\end{figure}

This strategy can also be used to reconstruct a smooth path between
two observed log-spectrograms $S_{ik}^{L_1}$ and $S_{ik'}^{L_2}$, in
this case the path being
\begin{equation}
\label{eq:match} S_{ik\rightarrow{ik'}}^{L_1\rightarrow L_2}(x)=[\mathfrak{F}_i^{x}]^{-1}
(x\mathfrak{F}_i^{L_1}(S_{ik}^{L_1})+(1-x)\mathfrak{F}_i^{L_2}(S_{i{k'}}^{L_2})),
\end{equation}
where a linear interpolation between the residuals takes the place
of the residual of the single language. This could be useful when it
is meaningful to pair two log-spectrograms in different languages,
for example because the same speaker is recorded in two languages.
This is not the case in our data set, but by way of example we
report in Fig.\ \ref{fig:sound_res} the path between the
log-spectrograms for the word \textit{un} for a French speaker
$S_{11}^{Fr}$ and the word \textit{um} for the Portuguese speaker who
is closest to the transformed $S_{11}^{Fr\rightarrow P}$. It is
also interesting to compare this with the interpolated path between
the two mean log-spectrograms in Fig.\ \ref{fig:sound_mean}.

Being able to extrapolate the sounds opens up interesting
possibilities whenever two languages are known to be at two stages of an
evolutionary path. In this case extrapolating in the direction of
the older (i.e. linguistically more conservative) language can provide an insight into the phonetic
characteristics of the extinct ancestor languages. This, of course,
will require some integration into a model of sound change with such information as that coming for example from textual analysis, history or
archaeology (e.g. dating studies) . This is also needed as the rate of change
of languages is not constant and the path $S^{L_1\rightarrow L_2}(x)$ can be travelled
at different speed for different branches of the language family's evolution,
and it can be changed by events such as conquests, migrations,
language contact, etc. However, by having a path in the first place, addressing such questions is now a possibility.

\subsection{Back to sound reproduction}

Visualizing the log-spectrograms (or other transformation of
the recorded sounds) is helpful but it is also important to listen to
the signals in the original domain. This is also true  for the
representation of a sound in a different language and the smooth
paths we have defined. Thus, we would like to reconstruct actual audible
sounds from the estimated log-spectrograms. To do this, we would
also need  information about the phase component that we have so far disregarded,
since we have focused all our attention on the amplitude component of the
Fourier Transform (see Section \ref{sec:data}). In principle, we
could perform a parallel analysis on the phases to obtain
representation of phase in a different language, the smooth path
between phases and so on. However, this is tricky from a
mathematical point of view, given the angular nature of the phases,
and in any case there is no reason to believe there is additional information captured by phase (human hearing is largely insensitive to phase so it is quite normal practice in acoustic phonetics to disregard the phase component). In practice, we use the phase
associated with the log-spectrogram $S_{ik}^{L_1}$ to reconstruct the
sounds over the smooth path; the results are quite
satisfactory. As supplementary material, some examples of reconstructed sound path can be provided.


\section{Discussion}\label{sec:disc}

We have introduced a novel way to explore phonetic differences and changes between
languages that takes into account the characteristics of the sound
population on the basis of actual speech recordings. The framework
we introduced is useful for dealing with acoustic phonetic data, i.e.
samples of sound recordings of different words or other linguistic units from
different groups (in our case, languages). We illustrate the
proposed method with an application to a Romance digit
data set, which includes the words corresponding to the numbers from
one to ten pronounced by speakers of five different Romance
languages. In particular, we verify in this data set the assumption
that the covariance structure in the log-spectrograms is common for
the different words within the language, thus increasing the sample
available for its estimation. This is an interesting example of how
the characteristics of a population (in this case the speakers of one
language) may be captured in the second order structure and not
only in the mean level. This in itself provides interesting information to linguists as it captures the notion of ``the sound of a language''. It also fits within the recent development of
object oriented data analysis \citep[see][]{Wang2007}, which
advocates a careful consideration of the object of
interest for a statistical analysis. Here it seems that marginal
covariance operators are promising features to represent
phonetic structure at the level of a language.

We do not focus here on the representativeness or otherwise of the
sample of speakers or words in the dataset. In view of a broad use of this approach
however, it is important to remember that the sample of speakers
should reflect the population we are interested in and in particular careful consideration should be given to regional and social stratification
in the data set. Moreover, to speak properly of a ``language'' (and not just of a small subset of words), the words considered should be
representative of the whole language. The digits studied here do contain a wide ranging set of different vowels and consonant (for just a few words), indicating that the results are likely to be generalisable to some extent across a larger corpus, but, of course, applying this to a much more comprehensive corpus of several languages would be advisable.

The proposed approach, using audio recordings in place of textual
representations, allows us to account for the differences between different
varieties of the same language, such as Castilian Spanish and American
Spanish \citep{penny2000}. Moreover, recent works \citep[see][and references
therein]{PhyloGroup2012,Bouchard2013,coleman2015} focus on the
reconstruction of the distribution of phonetic features for ancestor
languages. While the research in this field is still in its very earliest stages, as a better understanding of the historical evolution of
sounds becomes available, this can be integrated into our methods to
provide a reconstruction of how the speakers of extinct languages
might have sounded. The final goal is therefore to integrate the modelling of
the variability of speech within the language provided by our
approach with the dynamics of sound change established by
other researches both in linguistics and in statistics. We are confident that this will make a
substantial contribution to the ongoing project whose goal is audible reconstruction of words in the proto-languages.

We have illustrated the transformation of a speaker's speech from one language to another as a first example application in speech generation but other problems can be addressed in this framework. For example, the proposed approach to model sound processes can be
extended to take into account discrete or continuous covariates
associated to the mean and the covariance operators. These can be seen as function of the geographical coordinates or of
time-depth when studying dialects. While we treated the language
as a categorical variable, nothing prevents us seeing it as a
continuous process in space and time. Indeed, the definition of the
continuous path between two languages described in Section
\ref{sec:extra} can be seen as the first step in this direction,
since the abscissa $x$ of the path can be made dependent on external
variables. While we do not claim this can straightforwardly reproduce
the evolutionary branches in language history, it can still be a
useful starting point for more complex models.

The application of the proposed method is not
necessarily restricted to comparative linguistics. It can be useful
whenever a comparison between groups of sounds is needed, or indeed other complex wavelike signals. In the
future it will be interesting to explore micro-variation within a
language (dialects, spoken language in different subgroups of the
population) but also other types of sounds such as songs or even
sounds different from human speech, for example animal calls.

\section*{Supplementary Material}

Supplementary material is available on request from the corresponding author.

\section*{Acknowledgements}
John Coleman appreciates the support of UK Arts and Humanities Research Council grant AH/M002993/1, ``Ancient Sounds: mixing acoustic phonetics, statistics and comparative philology to bring speech back from the past''. John Aston appreciates the support of UK Engineering and Physical Sciences Research Council grant EP/K021672/2, ``Functional Object Data Analysis and its Applications''.

\bibliographystyle{natbib}

\end{document}